\documentclass[%
 preprint,
 amsmath,amssymb,
 aps, physrev,
]{revtex4-2}

\usepackage{graphicx}% Include figure files
\usepackage{dcolumn}% Align table columns on decimal point
\usepackage{bm}% bold math
\usepackage{romannum}
\usepackage{graphicx}% Include figure files
\usepackage{dcolumn}% Align table columns on decimal point
\usepackage{chemformula}

\usepackage{siunitx}
\usepackage[colorlinks, urlcolor=blue, citecolor=blue]{hyperref}
\hypersetup{
	colorlinks=true,
	linkcolor=blue,
	filecolor=blue,
	urlcolor=blue,
}
\usepackage{xcolor}
\definecolor{myred}{RGB}{255,3,0}
\definecolor{myblue}{RGB}{31,60,255}
\definecolor{myblue2}{RGB}{34,123,255}
\definecolor{myorange}{RGB}{226,48,0}
\definecolor{mygreen}{RGB}{21,100,0}
\definecolor{mygreen2}{RGB}{118,172,66}

\newcommand{\rev}[1]{{\color{black}\normalfont #1}}

\usepackage[normalem]{ulem}
\usepackage[textwidth=\dimexpr\textwidth-2cm\relax]{todonotes}
\makeatletter
\@mparswitchfalse%
\makeatother
\normalmarginpar %for right-handed notes and lines, or

\begin{document}

% \preprint{APS/123-QED}

\title{To jump or not to jump: Adhesion and viscous dissipation dictate the detachment  of coalescing wall-attached bubbles}

\author{\c{C}ayan Demirk{\i}r}
% \email[]{c.demirkir@utwente.nl}
\affiliation{
	Physics of Fluids Department, Max Planck Center Twente for Complex Fluid Dynamics, and J. M. Burgers Center for Fluid Dynamics, University of Twente, P.O. Box 217, 7500AE Enschede, Netherlands
}

\author{Rui Yang}
%\email[]{c.demirkir@utwente.nl}
\affiliation{
	Physics of Fluids Department, Max Planck Center Twente for Complex Fluid Dynamics, and J. M. Burgers Center for Fluid Dynamics, University of Twente, P.O. Box 217, 7500AE Enschede, Netherlands
}

\author{Aleksandr Bashkatov}
%\email[]{c.demirkir@utwente.nl}
\affiliation{
	Physics of Fluids Department, Max Planck Center Twente for Complex Fluid Dynamics, and J. M. Burgers Center for Fluid Dynamics, University of Twente, P.O. Box 217, 7500AE Enschede, Netherlands
}

\author{Vatsal Sanjay}
%\email[]{c.demirkir@utwente.nl}
\affiliation{
	Physics of Fluids Department, Max Planck Center Twente for Complex Fluid Dynamics, and J. M. Burgers Center for Fluid Dynamics, University of Twente, P.O. Box 217, 7500AE Enschede, Netherlands
}

\author{Detlef Lohse}%
% \email[]{d.lohse@utwente.nl}
\affiliation{
	Physics of Fluids Department, Max Planck Center Twente for Complex Fluid Dynamics, and J. M. Burgers Center for Fluid Dynamics, University of Twente, P.O. Box 217, 7500AE Enschede, Netherlands
}
\affiliation{
	Max Planck Institute for Dynamics and Self-Organisation, Am Fassberg 17, 37077 G{\"o}ttingen, Germany
}

\author{Dominik Krug}
%\email[Contact author: ]{d.j.krug@utwente.nl}
\email[Contact author: ]{d.krug@aia.rwth-aachen.de}
\affiliation{
	Physics of Fluids Department, Max Planck Center Twente for Complex Fluid Dynamics, and J. M. Burgers Center for Fluid Dynamics, University of Twente, P.O. Box 217, 7500AE Enschede, Netherlands
}

\affiliation{ Institute of Aerodynamics, RWTH Aachen University, Wüllnerstrasse 5a, 52062 Aachen, Germany
}

% \title{\textbf{The title should simply and concisely convey the main findings. Avoid nonstandard abbreviations and acronyms.} 
% }% 

% \author{Ann Author}
%  \altaffiliation[Also at ]{Physics Department, XYZ University.}%Lines break automatically or can be forced with \\
%  \email{Contact author: Second.Author@institution.edu}
% \affiliation{%
%  Authors' affiliations\\
%   Include all institutions where the work was conducted: department or division, institution, city, state (if relevant), and country, in this order.
% }%

%  \homepage{http://www.Second.institution.edu/~Charlie.Author}
% \affiliation{
%  First affiliation for this author
% }%
% \affiliation{
%  second institution for this author
% \affiliation{%
%  Authors' institution and/or address\\
%  This line break forced with \textbackslash\textbackslash
% }%

% \collaboration{CLEO Collaboration}%\noaffiliation

\date{\today}% It is always \today, today,

\begin{abstract}
Bubble coalescence can promote bubble departure at much smaller \rev{bubble sizes than those required for buoyancy-driven detachment.} This can critically enhance the efficiency of gas-evolving electrochemical processes, such as water electrolysis.
In this study, we integrate high-speed imaging experiments and direct numerical simulations to dissect how and under which conditions bubble coalescence on surfaces leads to detachment. Our transparent electrode experiments provide new insights into contact line dynamics, demonstrating that the bubble neck generally does not contact the surface during coalescence.
We reveal that whether coalescence leads to bubble departure or not is  determined by the balance between surface energy, adhesion forces, and viscous dissipation.
For the previously unexplored regime at low effective Ohnesorge number, a measure of viscosity that incorporates the effect of asymmetry between the coalescing bubbles, we identify a critical dimensionless adhesion energy threshold of $\approx$15\% of the released surface energy, below which bubbles typically detach. 
We develop a global energy balance model that successfully predicts coalescence outcomes across diverse experimental conditions. 
\end{abstract}

\maketitle

%\tableofcontents

% \section{\label{sec:level1}First-level heading:\protect\\ The line
% break was forced \lowercase{via} \textbackslash\textbackslash}

\section{\label{sec:intro}Introduction}

Coalescence events are ubiquitous and drive numerous natural and industrial phenomena, from gas exchange in aquatic ecosystems \cite{deike2022mass, jiang2024abyss} to raindrop formation \cite{low1982collision} and chemical reactors \cite{schluter2021small}. In particular, \rev{by initiating the departure of surface-attached bubbles at smaller sizes and through the release of surface energy rather than relying on buoyancy,} coalescence can affect the efficiency of many industrially relevant processes, such as boiling and gas-evolving chemical reactions \cite{xu2008pool, wang2014intensification}. In light of the ongoing energy transition, there is particular interest in the role of bubbles in electrochemical systems \cite{wang2009water, zhao2019gas}, where mitigating bubble effects is considered key to enabling cost-effective production of green hydrogen \cite{swiegers2021prospects}. Bubbles play a critical role in these systems, as they can block the active surface area, obstruct ion transport pathways and generally affect the mass transport \cite{angulo2020influence,sepahi2022effect}. Controlling  the bubble population via coalescence-induced detachment is attractive, since it enables earlier detachment compared to the buoyancy limits of the parent bubbles \cite{soto2018coalescence,iwata2022coalescing,lv2021self,park2024coalescence,raza2023coalescence} without external actuation. The concept has proven effective in increasing the performance in a micro-electrode model \cite{bashkatov2024performance}. However, applying this approach to practical system requires understanding the conditions under which coalescence triggers departure, which is the focus of this work. 
 
Coalescence of wall-attached bubbles can result in ``\textit{jumping}," where the merged bubble detaches, or ``\textit{sticking}," when it remains attached. During the event, surface energy is transformed into kinetic energy. The associated viscous dissipation differs fundamentally between droplets and bubbles coalescence. For droplets, viscous losses occur within the confined liquid volume with well-defined characteristic scales \cite{paulsen2014coalescence,vakarelski2024bubbles}. In contrast, for bubbles, dissipation occurs in the surrounding unbounded liquid, obscuring relevant length scales. \rev{Additionally, bubbles must overcome substantial liquid inertia \cite{soto2018coalescence,lv2021self,park2024coalescence} and adhesion energy \cite{park2024coalescence}, preventing direct application of droplet-based approaches, for which extensive literature exists \cite{liu2014numerical,wasserfall2017coalescence, yan2019droplet}, to bubbles.} \rev{Detachment criteria for bubbles so far rely on limited experimental data \cite{lv2021self,park2024coalescence}, with bubbles growing on pits and no information on the contact patch evolution, and numerical investigations \cite{iwata2022coalescing, zhao2022coalescence}, where the accurate representation of the contact line remains challenging. There is no consensus on the mechanism governing the jumping/sticking transition: Refs. \cite{iwata2022coalescing,zhao2022coalescence} propose a competition between neck touchdown and contact patch removal by capillary waves, while others favor energy arguments \cite{lv2021self,park2024coalescence}. For the latter, it remains unclear whether dissipation or adhesion energy dominates, and the  dissipation scaling proposed in \cite{lv2021self} is inconsistent with experimental data.}

In this paper, \rev{we address these open questions and} develop a criterion for bubble detachment after coalescence, incorporating material properties (density $\rho$, viscosity $\mu$, surface tension $\sigma$, equilibrium contact angle $\theta_\text{eq}$), and geometric parameters (bubble radius $R$, and contact patch radius $R_{\text{cont}}$). Through experiments and accompanying direct numerical simulations, we demonstrate that coalescence outcomes are primarily governed by the balance between available surface energy $\Delta G$, adhesion energy $W_{a}$, and viscous dissipation $W_\mu$.
Using a transparent electrode reveals, for the first time, contact line motion during bubble coalescence, enabling a comprehensive  energy balance analysis.
\vspace{-0.2cm}

\section{\label{sec:setup}Methods}

\subsection{\label{sec:level2_setup_exp}Experimental details}

\begin{figure*}[h!]
	\centering
	\includegraphics[width=0.8\textwidth]{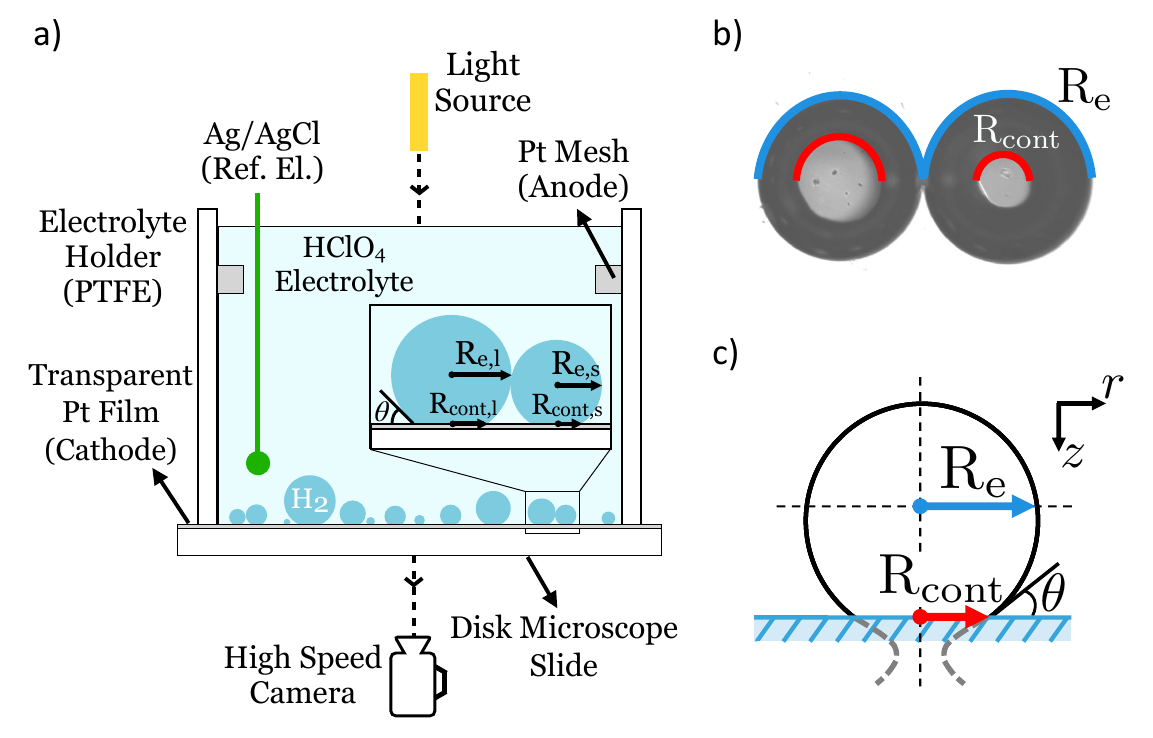}
	\caption{(a) Schematic of the shadowgraphy setup for observing the coalescence of wall-attached bubbles. (b) \rev{Typical experimental image of coalescing bubbles, used to determine $R_{cont}$ and $R_e$ (indicated by half-circles).}
    (c) \rev{Reconstructed bubble shape based on the Young-Laplace equation and the measured values of $R_{cont}$ and $R_e$.}
    }
	\label{fig:1_setup}
\end{figure*}

A custom-made electrolysis cell was used for the experiments, as outlined in Figure \ref{fig:1_setup}a. The cell featured a 40 mm diameter disk electrode \textcolor{black}{(cathode)}  comprising a 20 nm thin film of platinum (Pt) sputtered onto a microscope glass slide. A tantalum layer with 3 nm thickness applied to improve the adhesion between Pt and glass. A platinized titanium mesh served as the \textcolor{black}{anode}, while an Ag/AgCl electrode (in 3 M NaCl; BASi{\textsuperscript{\tiny\textregistered}}) functioned as the reference electrode. Further details on the cell and WE are given in previous works \cite{pande2020electrochemically, pande2021electroconvective, demirkir2024life}. The electrolyte consisted of 0.1 M and 1 M perchloric acid ($\mathrm{HClO_4}$) solutions. To improve the conductivity, 0.5 M sodium perchlorate monohydrate ($\mathrm{NaClO_4.H_20}$) was added to the 0.1 M solution as a supporting electrolyte. The chemicals used in the experiments were supplied from Sigma-Aldrich (purity of 99.99\%). A Biologic VSP-300 potentiostat was used to perform the chronopotentiometry experiments. The applied current densities ($j$) are kept low ($\lvert j \rvert \leq200\:\mathrm{A/m^2}$), such that potential thermal \cite{massing2019thermocapillary} or solutal \cite{park2023solutal} Marangoni effects or electrostatic forces remain negligible \cite{demirkir2024life}.

The thin layer of platinum deposited on the glass substrate provides optical transparency to the working electrode, allowing for high-resolution visualization and analysis of the bubbles from underneath (see Fig. \ref{fig:1_setup}a). Bubbles were monitored by a high-speed camera (Photron Nova S12) with frame rates between 10 Hz and 50,000 Hz, employing a 5$\times$ magnification objective lens (Olympus MPLFLN) and backlight illumination. The calibration procedure was explained in a previous work \cite{demirkir2024life}. From the resulting images \rev{(see Fig. \ref{fig:1_setup}(b))}, we extract the equatorial radii $R_{e, i}$ (with subscript $i = l,s$ denoting the large ($l$) or the small ($s$) bubble) based on
the black regions and the contact patch radii $R_{\text{cont}, i}$ from the central gray areas. Bubbles with identical $R_{e, i}$ can differ in volume due to variations in $R_{\text{cont}, i}$. Therefore, we use the volume equivalent radius $R_i$ to characterize the bubble size. To obtain $R_i$ \rev{along with the contact angle $\theta$}, we solve the Young--Laplace equations \cite{chesters1978modes} for the bubble shape \rev{(see Fig.~\ref{fig:1_setup}(c))}, assuming rotational symmetry with $R_{\text{cont}, i}$ and $R_{e, i}$ as constraints \cite{demirkir2024life}.
\rev{These shapes also provide the initial conditions for our volume-of-fluid based direct numerical simulations (DNS), performed using the open-source Basilisk C language \cite{popinet2009accurate,basilliskpopinet,Sanjay2024code}.}

\subsection{\label{sec:level2_setup_num}Governing equations and numerical setup}

To further elucidate the dynamics of coalescence-induced bubble jumping, we use the volume of fluid (VoF) based finite volume method implemented in the open-source software Basilisk C \cite{popinet2009accurate,basilliskpopinet}. This approach solves the mass and momentum conservation equations given by
\begin{align}
	\boldsymbol{\tilde{\nabla}\cdot\tilde{v}} & =0, \\
	\frac{\partial \boldsymbol{\tilde{v}}}{\partial \tilde{t}} + \boldsymbol{\tilde{\nabla}\cdot}\left(\boldsymbol{\tilde{v}\tilde{v}}\right) & =\frac{1}{\tilde{\rho^*}}\left(-\boldsymbol{\tilde{\nabla}} \tilde{p}'+ \boldsymbol{\tilde{\nabla}} \cdot\left(2 Oh^* \boldsymbol{\tilde{\mathcal{D}}}\right)+\boldsymbol{f}_\delta\right),
\end{align}

\noindent where lengths are normalized by the pre-coalescence single bubble radius. The velocity $\boldsymbol{v}$ and pressure $p$ are normalized using the inertio-capillary velocity $U_{\sigma}=\sqrt{\sigma / \rho_l R}$ and capillary pressure $P_\sigma=\sigma / R$, respectively. Note that, in the simulations in this letter, we only focus on the symmetric bubbles $R_l = R_s = R$.
Here, $\boldsymbol{\mathcal{D}}$ represents the symmetric part of the velocity gradient tensor, and $\boldsymbol{f}_\delta$ is the singular body force at the liquid-gas interface. Following \cite{popinet2018numerical}, we redefine pressure as $\tilde{p}^{\prime}=\tilde{p}+Bo\tilde{\rho}\tilde{z}$, where the Bond number, $Bo=\rho g {R}^2/\sigma$, compares gravitational and capillary pressures. The singular body force is given by:
\begin{align}
	\boldsymbol{f}_\delta \approx \left(\tilde{\kappa} + Bo\left(1-\frac{\rho_\text{gas}}{\rho}\right)\tilde{z}\right)\boldsymbol{\tilde{\nabla}}\Psi,
\end{align}

\noindent with $\Psi$ as the volume of fluid (VoF) color function distinguishing liquid ($\Psi = 1$) from gas ($\Psi = 0$). Using the one-fluid approximation \cite{tryggvason2011direct}, we express the Ohnesorge number $Oh^*$ and dimensionless density $\tilde{\rho^*}$ as:
\begin{align}
	Oh^* & = Oh\left(\Psi +(1-\Psi)\Bigl(\frac{\mu_\text{gas}}{\mu}\Bigr)\right), \\
	\tilde{\rho^*} & =\Psi+(1-\Psi) \Bigl(\frac{\rho_\text{gas}}{\rho}\Bigr),
\end{align}

\noindent where $\mu_g/\mu$ and $\rho_g/\rho$ are gas-liquid viscosity and density ratios, respectively. The liquid Ohnesorge number is defined as:
\begin{equation}
	Oh=\frac{\mu}{\sqrt{\rho \sigma R}}
\end{equation}

To streamline our investigation, we fix $\rho_\text{gas}/\rho$ at $7.82\times10^{-5}$ and $\mu_\text{gas}/\mu$ at $8.8\times10^{-3}$, reflecting realistic hydrogen properties. We maintain $Oh$ at $7.5\times10^{-3}$ for all cases except those directly comparing with experiments. Furthermore, we employ the simplest contact line model where we prescribe the contact angle $\theta_{eq}$ in the first computational cell adjacent to the electrode. A numerical slip regularizes the well-known contact line singularity \cite{afkhami2018transition, snoeijer2013moving}. For detailed methodology and the Basilisk C codes to solve the above equations, we refer readers to \cite{Sanjay2024code}.

\section{\label{sec:results}Results and Discussion}

Based on the bubble parameters along with the material properties of the electrolyte and hydrogen, we can estimate the released surface energy $\Delta{G} \sim \sigma (R_l^2+R_s^2-R_m^2)$
\noindent where $R_m = \left(R_l^3+R_s^3\right)^{1/3}$ is the volume equivalent radius of the merged bubble, and the total adhesion energy
$W_{a,\text{tot}} \sim \sigma \cos{\theta_{\text{eq}}}(R_{cont,l}^2+R_{cont,s}^2)$,
\noindent based on which we define the dimensionless adhesion energy
\begin{align}
	W_{a,\text{tot}}^* \equiv \frac{\cos{\theta_{eq}}(R_{cont,l}^2+R_{cont,s}^2)}{R_l^2+R_s^2-R_m^2},
	\label{eq:nonDimAdhesionEnergy}
\end{align}
as a control parameter. Here, the equilibrium contact angle $\theta_{\text{eq}}$ characterizes the surface wettability, which directly determines the adhesion force between the bubble and the solid surface. Additionally, the system features three more dimensionless control parameters, namely the Ohnesorge number $Oh $ (dimensionless electrolyte viscosity), the radius ratio $x \equiv R_l/R_s$, and the Bond number $Bo$ (dimensionless gravity). 

\subsection{\label{sec:level2_x=1}Key observations for $x \approx 1$}

\begin{figure*}[h!]
	\centering
	\includegraphics[width=\textwidth]{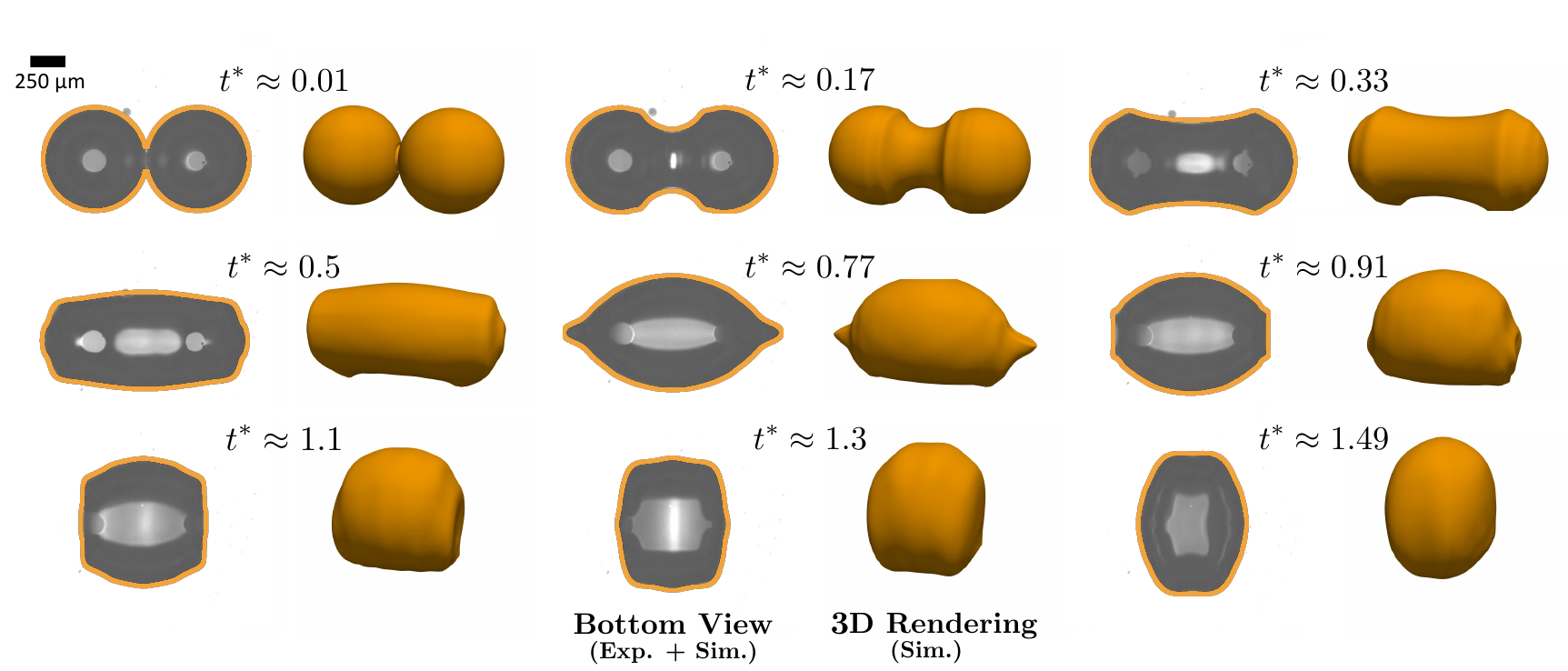}
	\caption{Shape and contact line evolution of two coalescing bubbles on a surface comparing experiment and simulations. Each pair shows: (left) experimental bottom view (grayscale) overlaid with numerical simulation contour (orange), and (right) 3D rendering from numerical simulation. In the experiments: $R_l \approx \SI{323}{\micro\meter},  R_s \approx \SI{321}{\micro\meter}$, $R_{\text{cont},l} = \SI{59}{\micro\meter}$ and $R_{\text{cont},s} = \SI{72}{\micro\meter}$. In the simulations: $R = \SI{322}{\micro\meter}$ and $R_{\text{cont}} = \SI{59}{\micro\meter}$. $Oh = 0.0066$, $Bo = 0.016$, {and $W^*_{a,\text{tot}} \approx 0.07$.} See the full movie in Suppl. Mat. }
	\label{fig:2_snapshots}
\end{figure*}

Figure~\ref{fig:2_snapshots} demonstrates the coalescence of two similarly sized bubbles ($R_l \approx R_s = R$) through experimental bottom-view grayscale images and numerical simulations. Time $t^*= t/\tau$ is normalized using the inertio-capillary timescale $\tau = \sqrt{{{\rho}{R^3}}/{\sigma}}$ \cite{eggers2024coalescence}. The numerical results for two identical bubbles, represented by orange outlines overlaid on the experimental images and three-dimensional renderings, show remarkable agreement with the experiments. This confirms that potential additional factors, such as contact line hysteresis, electric forces, and Marangoni stresses---neglected in the simulations---are of only secondary importance in the present problem. As the bubbles merge, a neck forms between them, expanding rapidly due to the strong Laplace pressure ($\Delta P\sim\sigma R/r_n^2$), where $r_n$ is the neck radius \cite{paulsen2014coalescence, oratis2023coalescence, raza2023coalescence}. This expansion follows the established scaling $r_n \sim t^{1/2}$ (not shown), resulting from the hydrodynamic singularity at the intersection of two spherical bubbles \cite{eggers2024coalescence}.
Initially, the neck expands symmetrically in the radial direction (Fig.~\ref{fig:2_snapshots}, $t^* \approx 0.01$-$0.33$). Subsequently, the presence of the wall breaks this symmetry (Fig.~\ref{fig:2_snapshots}, $t^* \approx 0.5$, see also \cite{soto2018coalescence}). Importantly, the lower neck never contacts the wall during coalescence, presumably due to the strong lubrication pressure between the neck and the surface. The experimental images clearly show the contact lines of both bubbles, confirming the presence of this gap. This holds true irrespective of whether the coalescence event results in detachment (as is the case in Fig.~\ref{fig:1_setup}b, or not (see Figs.~\ref{fig:SM_snapshot_sticking} and \textcolor{blue}{S1})). This scenario is at odds with the simulation results of \cite{iwata2022coalescing}, who identified the touchdown of the neck and the resulting enlargement of the contact patch as the key effect resisting bubble departure. 

Following the neck expansion, \rev{capillary waves emerge}
\rev{and propagate along the bubble surfaces}, significantly altering their shape. For (nearly) identically sized bubbles, waves converge at the two apices simultaneously, generating a characteristic lemon-like shape at $t^* \approx 0.77$ (Fig.~\ref{fig:2_snapshots}). Subsequently, surface tension drives a retraction wave inwards (Fig.~\ref{fig:2_snapshots}, $t^* \approx 0.91$), initiating contact line motion from the outer edge at $t^* \approx 1.1$. The presence of contact angle hysteresis in the experiments ($\approx 70^\circ$ despite low mean roughness $\approx 1$ nm \cite{demirkir2024life}) ensures that the contact line of both bubbles remains pinned until the waves return to the contact line \cite{mittal2003contact, hong2011anomalous}. When capillary waves reach the contact line, either the entire contact patch is swept away, leading to bubble detachment (as observed at $t^* \approx 1.49$ in Fig.~\ref{fig:2_snapshots}), or the bubble remains stuck to the surface. In the latter case, the smaller of the two contact patches typically still gets peeled off, but the larger patch remains attached retaining the bubble. This is particularly evident from Fig.~\ref{fig:8_cont_area}, which shows that the contact area after coalescence ($A_{cont,m}$) is usually close to the one of the larger patch ($A_{{cont},lp}$) before the event, especially as $x\gtrapprox1.8$. Note that the larger patch, denoted with subscript 'lp', is typically but not exclusively associated with the larger bubble. Persistent shape oscillations after coalescence modulate the contact patch size of the sticking bubble after the event in some instances. Only for rare cases at $x \approx 1$, we observe a ``moving contact patch", where both contact areas converge and merge at the bubbles' center (see section \textcolor{blue}{S1}). 

\begin{figure}[t!]
	\centering
	\includegraphics[width=0.8\columnwidth]{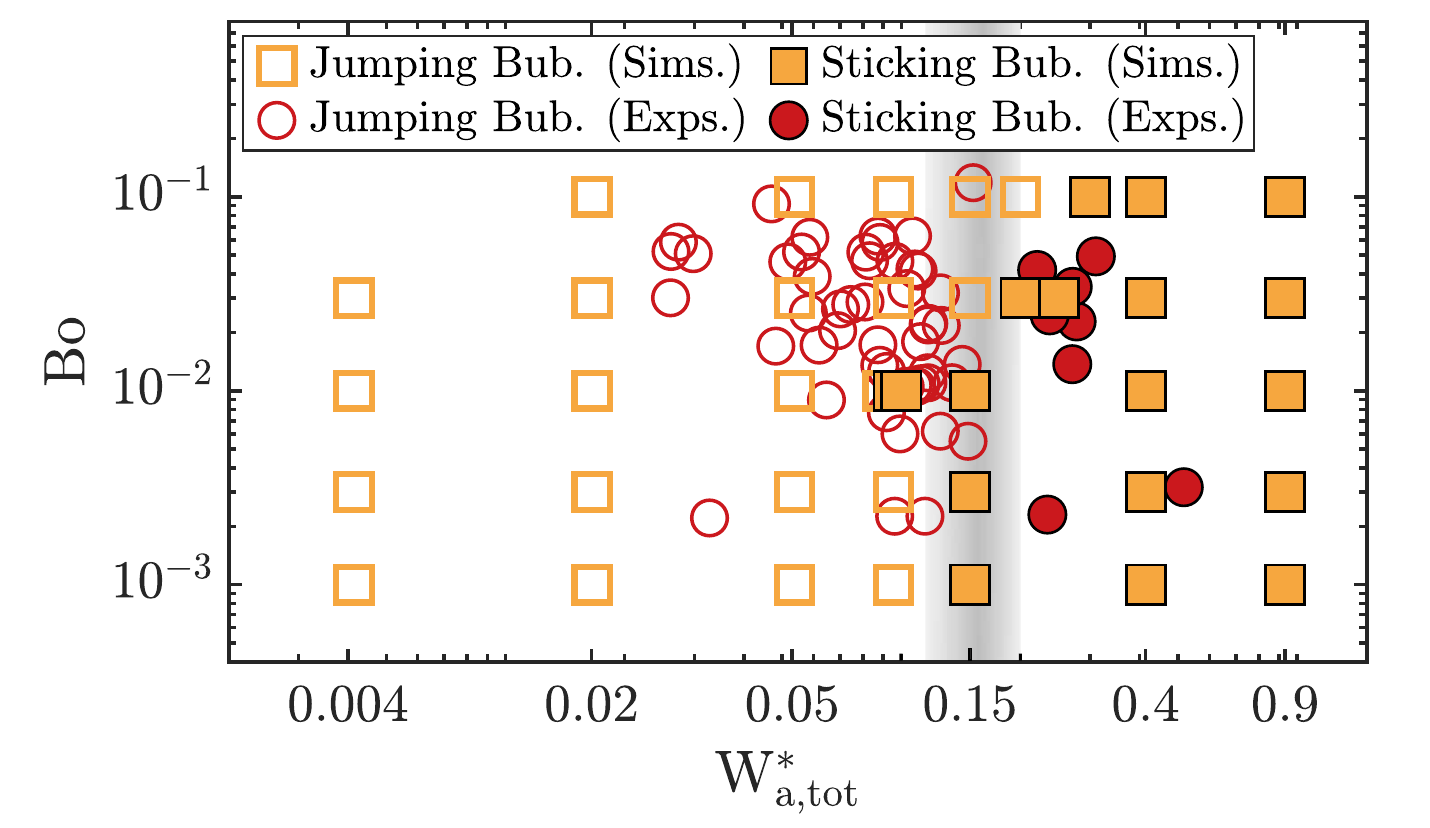}
	\caption{Coalescence outcomes for similarly sized bubbles ($x < 1.2$ and $\mathrm{R_{cont,l}/R_{cont,s} < 1.2}$) in the experiments (red), and identical ones the simulations (orange) across the Bond number $Bo$ vs. dimensionless adhesion energy $W^*_{a, \text{tot}}$ phase space. Open and filled markers denote jumping and sticking cases, respectively.
    Gray shading marks the jumping-sticking transition region.}
	\label{fig:2_Bo_Wa}
\end{figure}

Figure~\ref{fig:2_Bo_Wa} summarizes the coalescence outcomes for the symmetrical case with $x \approx 1$ across a range of $Bo$ and $W^*_{a, \text{tot}}$ values, comparing simulations (orange squares) and experiments (red circles). Empty markers denote jumping, while filled markers indicate sticking. Notably, the outcome exhibits only a weak dependence on $Bo$ over two orders of magnitude of $Bo$ ($10^{-3} < Bo < 10^{-1}$), with $W^*_{a,\text{tot}}$ emerging as the critical parameter governing the jumping-sticking transition in this regime. This transition occurs around  $W^*_{a,\text{tot}} \approx 0.15$ (indicated by the grey shaded area in the plot), implying that bubbles fail to jump when adhesion energy exceeds approximately 15\% of the total released surface energy (at $Oh \approx 0.0075$). The numerical results are largely consistent with this threshold despite the lack of contact line hysteresis in the simulations, indicating that this is a secondary effect. The slight trend in the jumping threshold towards higher $W^*_{a,\text{tot}}$ at the upper $Bo$ limit visible for the simulations may results from increased buoyancy forces favoring detachment. Overall, these results clearly demonstrate the important role the adhesion energy plays in the problem. Taking this into account, we will now extend our analysis to asymmetric coalescence events with size ratios $x>1$.

\subsection{\label{sec:level2_x>1}Adhesion limited jumping inhibition of asymmetric bubbles}

\begin{figure}[h!]
	\centering
	\includegraphics[width=0.9\columnwidth]{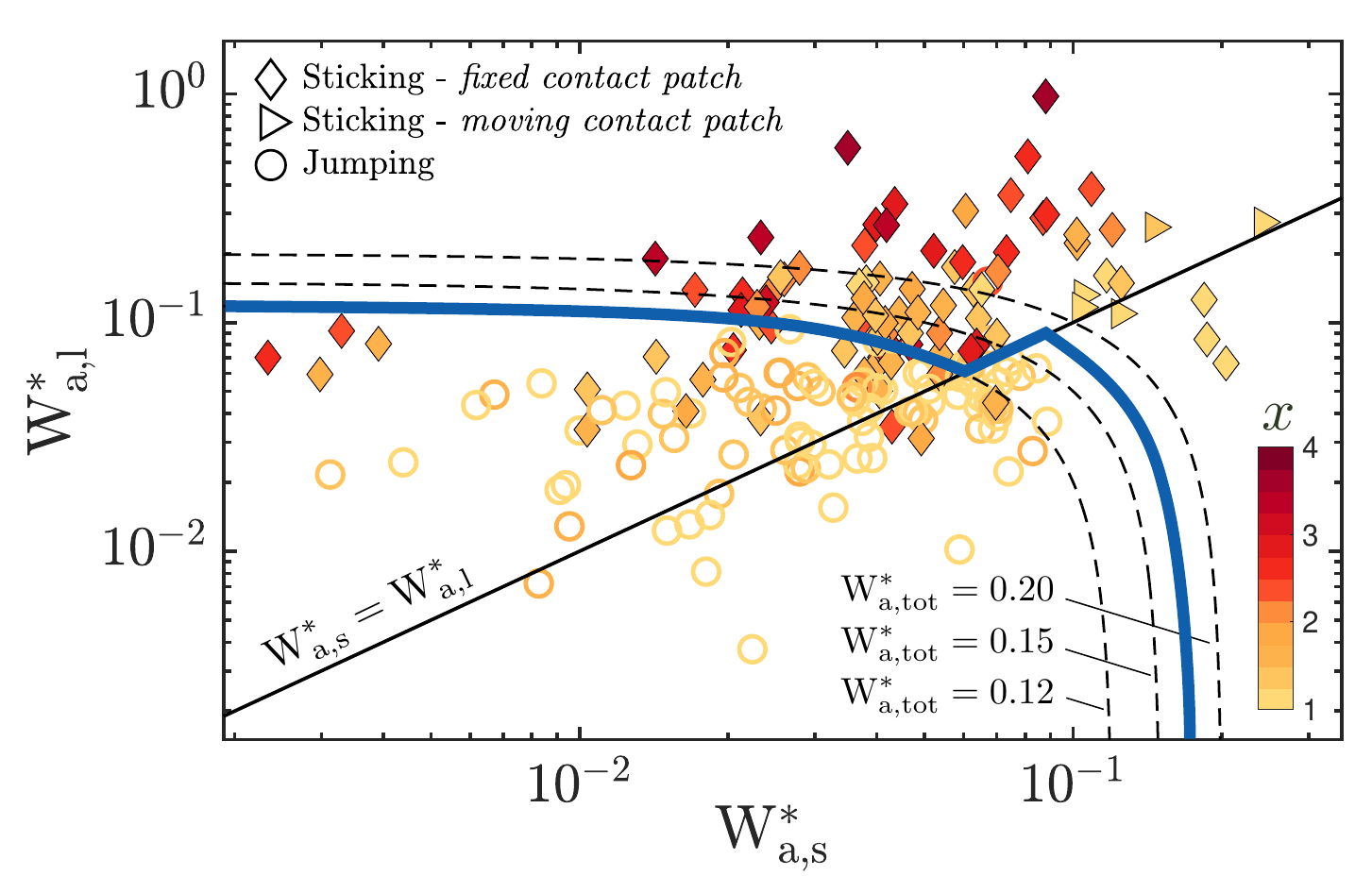}
	\caption{Sticking (filled) and jumping (open markers) as a function of the dimensionless adhesion energies of the larger ($W_{a,l}^*$) and smaller ($W_{a,s}^*$) bubbles. The diagonal line represents $W_{a,l}^* = W_{a,s}^*$, dashed lines are isocontours of the total dimensionless adhesion energy $W_{a,\text{tot}}^*$.}
	\label{fig:4_Wa_LargeSmall}
\end{figure}

% {\it Adhesion limited jumping inhibition of asymmetric bubbles}: 
For asymmetric bubbles, the total dimensionless adhesion energy $W^*_{a,\text{tot}}$, but also its distribution between the large ($W^*_{a,l}$) and small ($W^*_{a,s}$) bubbles may influence the coalescence process. Figure~\ref{fig:4_Wa_LargeSmall} presents our experimental results in the $W^*_{a,l}$ -- $W^*_{a,s}$ phase space, classifying jumping and sticking bubbles. Each data point represents a coalescence event, with marker style indicating the outcome (filled symbols for sticking, open for jumping) and shade representing the bubble size ratio $x$. The data discussed up to now falls close to the diagonal line,  denoting equal distribution of the adhesion energy between the bubbles, while dashed lines indicate isocontours of $W^*_{a,\text{tot}}$. Cases where the merged bubble is expected to detach due to buoyancy are omitted from the analysis \rev{(see section S2 of \cite{supplMaterial})}.

As expected, larger bubbles typically exhibit higher adhesion energy than smaller ones, which is reflected in the concentration of points above the diagonal. For these ``regular" cases, the adhesion energy threshold for jumping inhibition $W_{a,\text{tot}}^* \approx 0.12$ is similar to that of identical bubbles ($W_{a,\text{tot}}^* \approx 0.15$). However, ``irregular" cases, for which the smaller bubble has a higher adhesion energy, show jumping inhibition only at slightly higher thresholds ($W_{a,\text{tot}}^* \approx 0.15$-$0.20$), indicating that this configuration favors detachment. The blue solid line approximates this transition between jumping and sticking. Furthermore, a significant correlation exists between the bubble size ratio $x$ and coalescence outcome, with larger ratios increasing the probability of sticking. The vast majority of jumping events occur for $x \lessapprox 1.75$ (\rev{Fig. S3 in}  \cite{supplMaterial}). This observation aligns with the fact that coalescence of similarly-sized bubbles releases more energy than that of bubbles with the same total volume but greater size disparity \cite{lv2021self,bashkatov2024electrolyte,sanjay2024asymmetry}.

\subsection{\label{sec:level2_global_en}Global energy balance}

The fate of the merged bubble--whether it jumps or remains stuck--can be predicted by the global energy balance in a coalescence event. The surface energy released during bubble coalescence $\Delta{G}$, resulting from the reduced gas-liquid interfacial area, is transformed into: energy overcoming adhesion forces $W_{a,\text{tot}}$, viscous dissipation $W_{\mu}$, change in potential energy $\Delta{E_{pot}}$, and translational kinetic energy $E_{kin}$ if the merged bubble detaches. $\Delta{E_{pot}}$ results from the slight shift in the center of mass of the bubbles before and after coalescence and is negligible compared to $\Delta{G}$. $E_{kin}$ constitutes a minor fraction of the total released energy across all investigated cases (see Appendix~\textcolor{red}{B}). The jumping-sticking transition is defined by $E_{kin} \approx 0$. Consequently, the global energy balance for this case yields,
\begin{align}
	\Delta{G} = W_{a,\text{tot}} + W_{\mu}.
	\label{eq:energy_bal}
\end{align}

\noindent For the viscous dissipation due to the velocity gradients generated in the surrounding liquid during coalescence, we derive the scaling  
\begin{equation}
	W_{\mu} \sim  \mu_\text{gas} \sqrt{\sigma/\rho} R_m^{3/2},
	\label{eq:dissipationScale}
\end{equation}
with details provided in Appendix~\textcolor{red}{C}.
We emphasize that the dissipation primarily occurs in the liquid and $\mu_\text{gas}$ only enters in Eq. (\ref{eq:dissipationScale}) through the interface continuity condition. 
\begin{figure}[b]
	\centering
	\includegraphics[width=\columnwidth]{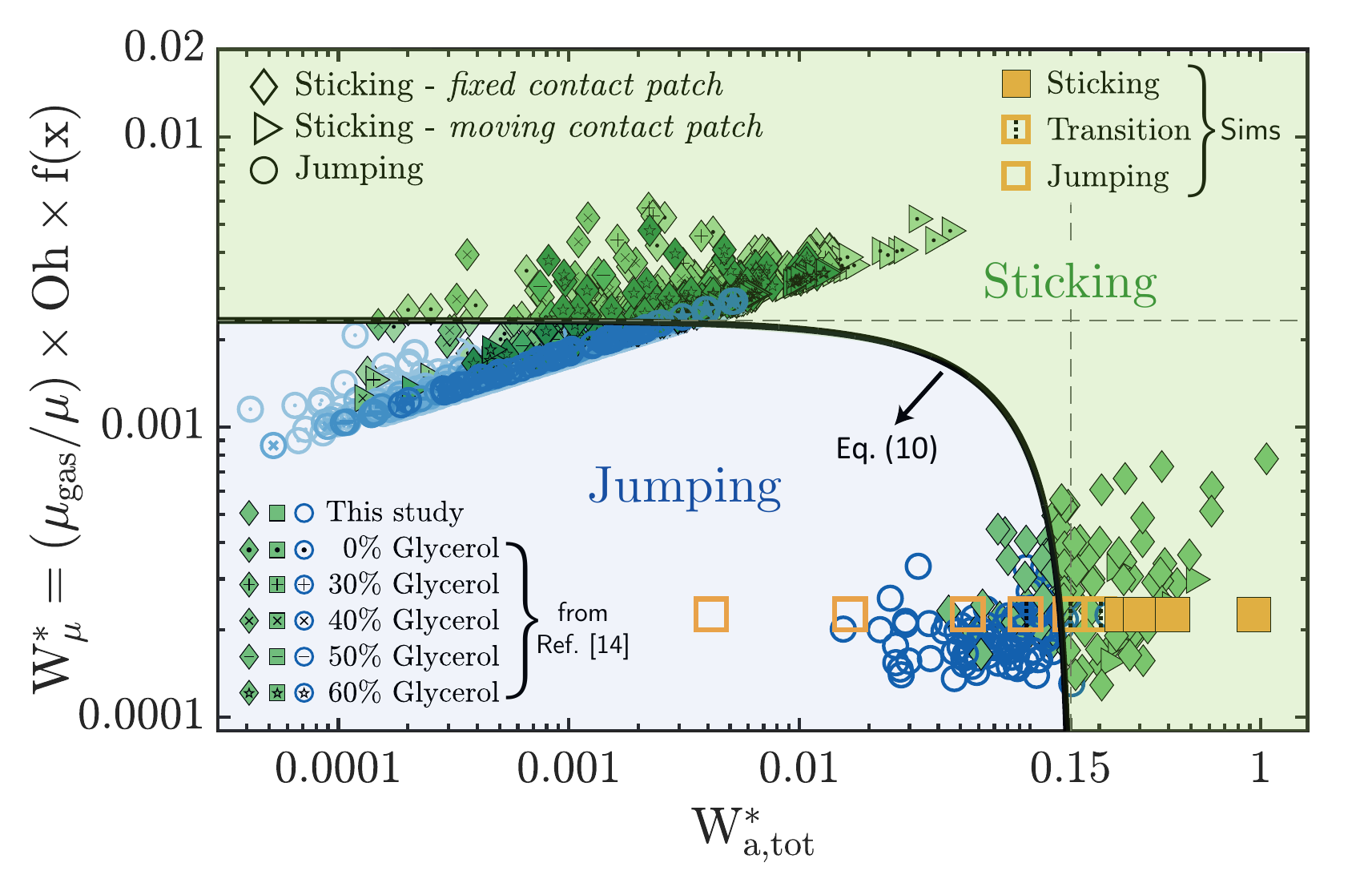}
	\caption{Generalized detachment prediction model for coalescing wall-attached bubbles. Data from two different experimental setups are included: small bubbles (up to $\approx$150 ${\mu}m$) with pinned contact line (upper left part, taken from Ref. \cite{lv2021self}) and larger bubbles (up to $\approx$1250 ${\mu}m$) with spreading contact line (lower right part, our experiments). Simulation results correspond to those in Fig.~\ref{fig:2_Bo_Wa}. The black curve depicts the best fit threshold curve base on Eq.~\eqref{eq:energy_bal_nonDim_star}. The asymptote values, $W_{a,\text{tot}}^* \approx 0.150$ and $W_{\mu}^* \approx 0.00233$ are shown as vertical and horizontal dashed lines, respectively. \rev{See section S4 of \cite{supplMaterial} for the details of parameters investigated in the plot.}}
	\label{fig:4_Wmu_W_a}
\end{figure}
\noindent Inserting the respective scaling relations for the terms in Eq.~\eqref{eq:energy_bal} and normalizing by 
$\Delta{G}$ results in 
\begin{equation}
	\alpha_1{W_{a,\text{tot}}^*}+\alpha_2{W_{\mu}^*}=1 ,
	\label{eq:energy_bal_nonDim_star}
\end{equation}

\noindent with the dimensionless adhesion energy $W_{a,\text{tot}}^*$ (see Eq.~\eqref{eq:nonDimAdhesionEnergy}) and the normalized viscous dissipation 
\begin{equation}
    W_{\mu}^* \equiv \frac{\mu_{gas} \sqrt{\sigma/\rho_l} R_m^{3/2}}{ \sigma(R_l^2+R_s^2-R_m^2)}.
\end{equation}
% $W_{\mu}^* \equiv (\mu_{gas} \sqrt{\sigma/\rho_l} R_m^{3/2})/( \sigma(R_l^2+R_s^2-R_m^2))$. 
The parameters $\alpha_1$ and $\alpha_2$ account for the transition from exact balance to scaling relations. Volume conservation during coalescence implies $R_m = (1+x^3)^{1/3} R_s$, allowing $W_{\mu}^*$ to be expressed as an effective Ohnesorge number $W_{\mu}^* = (\mu_{gas}/\mu)\cdot Oh \cdot f(x)$, where
\begin{align}
	f(x) = \frac{\sqrt{x(1+x^3)}} {x^2 + 1 - (1+x^3)^{2/3}}
	\label{eq:f_x}
\end{align}

\noindent  depends solely on the radius ratio $x$, which acts to increase the effective Ohnesorge number for asymmetric coalescence events (see Fig. \ref{fig:fx_change}).

Figure~\ref{fig:4_Wmu_W_a} presents the data in the $W_{a,\text{tot}}^*$-- $W_{\mu}^*$ parameter space of Eq. (\ref{eq:energy_bal_nonDim_star}). To complement the present dataset, we also include data from the experiments by \citet{lv2021self} here. In their case, oxygen bubbles are formed by a catalytic reaction in $\mathrm{H_2O_2}$ solution on an Au surface with contact lines pinned at predefined pits. The pits have radii of \SI{1}{\micro\meter}, which is the value assumed as contact radius for these cases.

Varying amounts of glycerol are added to change the liquid parameters and in particular the viscosity $\mu$, which is increased more than tenfold at the highest concentration. The resulting bubbles are significantly smaller ($\approx$10 times in size and $\approx$100 times in contact patch) than those in our experiments, which provides access to an entirely different region of the parameter space. To determine the prefactors $\alpha_1$ and $\alpha_2$ in Eq.~\eqref{eq:energy_bal_nonDim_star}, we seek the parameter combination that yields the largest share of correct attributions of `jumping' and `sticking' cases for both datasets. Doing so results in  $\alpha_1 = 6.66 $ and $\alpha_2 = 429.2$, for which $\approx$88\% of the total number of events is predicted successfully. The corresponding jumping-sticking transition curve  based on Eq.~\eqref{eq:energy_bal_nonDim_star} is included as black line in Fig. \ref{fig:4_Wmu_W_a}. 

In the limit of low adhesion energy, where most of the data of \cite{lv2021self} lies, the outcome is solely determined by dissipation and the jumping threshold in the asymptotic limit is given by $W_{\mu}^* \approx 0.00233$. In contrast, the present data falls towards the limit of small $W_{\mu}^*$, where the fate of the bubble is determined by adhesion with a critical value of $W_{a,\text{tot}}^* \approx 0.150$ in the limit of negligible dissipation. 

\section{\label{sec:results}Conclusions}

In conclusion, we have shown that in addition to the viscous dissipation also the adhesion energy of surface-attached coalescing bubbles plays a critical role in determining their fate. This is based on the direct experimental observation of the contact line dynamics during coalescence and consistent with numerical simulations at matched conditions. By evaluating the energy balance at the sticking-jumping transition, we derived a detachment criterion that captures and is consistent with all available data on this problem. One of the two relevant input parameters is an effective Ohnesorge number, that accounts for size disparity and the gas-liquid viscosity ratio. The other parameter is the dimensionless adhesion energy, which can be related to the contact angle when evaluating the criterion for a particular bubble configuration. The developed expression has strong support with data spanning several orders of magnitude in both parameters, and can be used e.g. to optimize surface structures of electrodes or boilers to enhance bubble release. \rev{Note that the criterion developed here features no dependence on electrolyte properties other than $\theta_{eq}$. At least for the numerically accessible parameter range, this is supported by the agreement with the simulations, for which potential electrolyte dependent effects such as Marangoni stresses are absent.}

\begin{acknowledgments}
We thank Pengyu Lv for providing the original data in his study. We also thank Andrea Prosperetti and Wilko Rohlfs for stimulating discussions.
This research was supported by the Dutch Research Council (NWO) through the ENW PPP Fund for top sectors, with contributions from Shell, Nobian, and Nouryon and the Ministry of Economic Affairs via the PPS-toeslagregeling, Grant No. 741.019.201. Additional funding was provided by the FIP-II project, sponsored by NWO and Canon, and the European Research Council (ERC) under the European Union's Horizon 2020 research and innovation programme (grant agreement No. 950111, BU-PACT). We also acknowledge support from the ERC Advanced Grant MultiMelt (No. 101094492). The numerical simulations were carried out on the national e-infrastructure of SURFsara, a subsidiary of SURF cooperation, the collaborative ICT organization for Dutch education and research.
\end{acknowledgments}

% \pagebreak

\rev{\section*{Data availability}

The data reported in this paper are available upon request.
}
\appendix

\section{\label{app:appendix_A}Sticking bubbles}
\setcounter{figure}{0}
\renewcommand{\thefigure}{A\arabic{figure}}
\setcounter{equation}{0}
\renewcommand{\theequation}{A\arabic{equation}}

\begin{figure}[t!]
	\centering\includegraphics[width=\columnwidth]{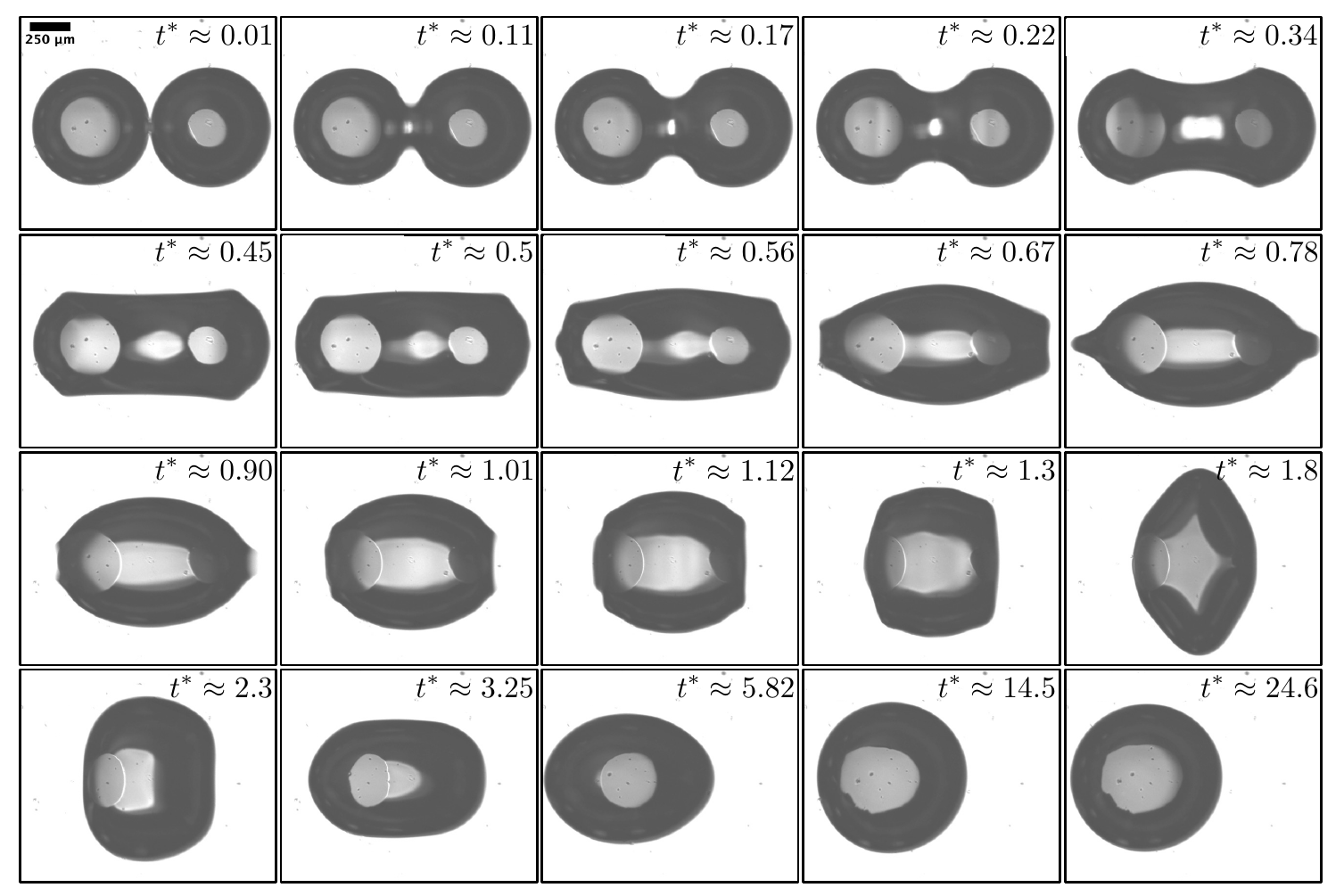}
	\caption{Outline and contact patch of coalescing bubbles on a planar electrode recorded at 20,000 frames per second. The merged bubble keeps sticking at the electrode surface after coalescence. The parameters are: $R_l \approx \SI{371}{\micro\meter},  R_s \approx \SI{360}{\micro\meter}$, $R_{\text{cont},l} = \SI{118}{\micro\meter}$ and $R_{\text{cont},s} = \SI{189}{\micro\meter}$. $Oh = 0.0062$, $Bo = 0.021$, {and $W^*_{a,\text{tot}} \approx 0.31$. }See the full movie in Suppl. Mat. \cite{supplMaterial}}
	\label{fig:SM_snapshot_sticking}
\end{figure}

The shape deformation and contact line motion during the coalescence of two bubbles of similar size, whose coalescence results in sticking, are shown in Fig. \ref{fig:SM_snapshot_sticking}. Similar to the coalescence-induced jumping shown in Fig. \ref{fig:2_snapshots} the neck formed during coalescence does not touch the electrode surface throughout the process. The smaller (right) contact patch becomes detached after $t^* \approx 1$, while the larger (left) patch, and consequently the merged bubble, remain attached. 

\begin{figure}[ht!]
	\centering
	\includegraphics[width=0.8\columnwidth]{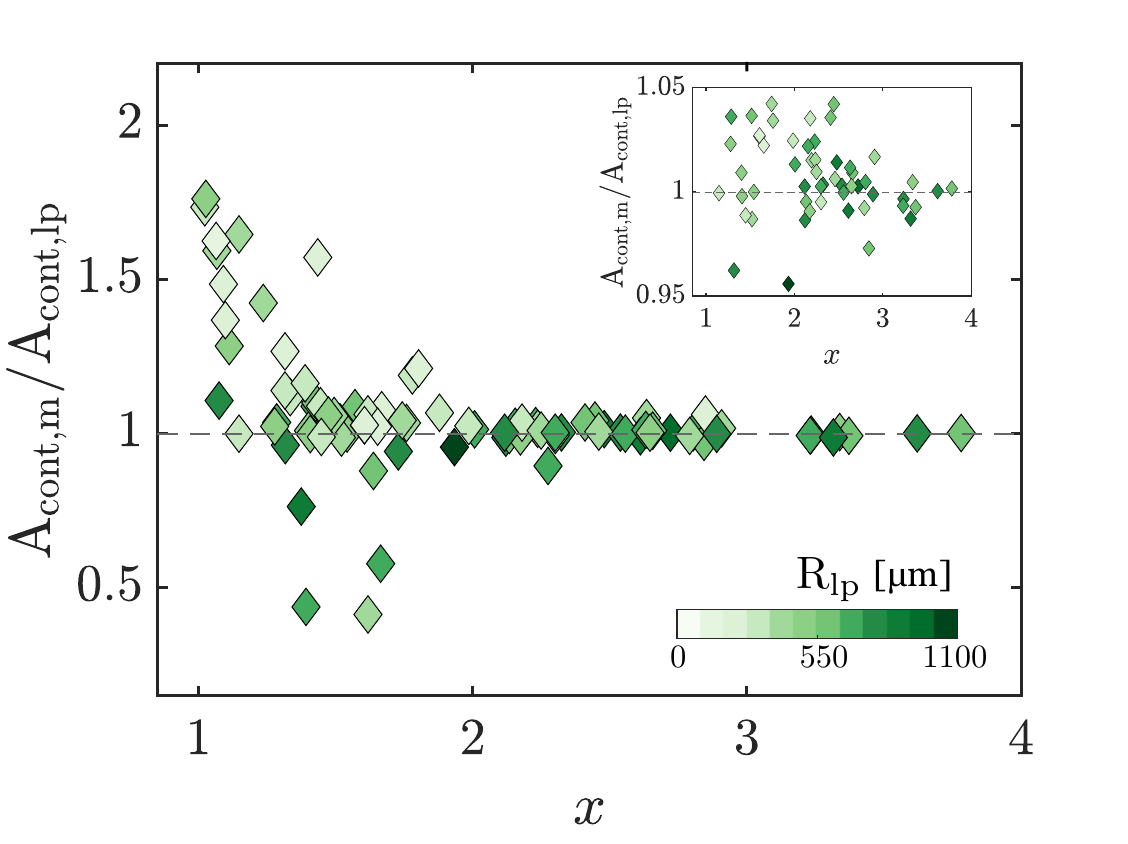}
	\caption{Change in contact patch area of the bubble with larger contact patch ($\mathrm{A_{cont,m} /A_{cont,lp}}$) with respect to the radius ratio of the coalescing bubbles ($x$). Color code indicates the size of bubble with larger patch patch ($\mathrm{R_{lp}}$). The inset shows a zoomed view in the vicinity of $\mathrm{A_{cont,m} /A_{cont,lp}} = 1$. }
	\label{fig:8_cont_area}
\end{figure}

Figure \ref{fig:8_cont_area} illustrates how the contact area of the fixed bubbles changes due to the coalescence events, with respect to the size ratio $x$ of the coalescing bubbles. The relative change in contact area is expressed as the ratio of the merged bubble's contact area after coalescence ($\mathrm{A_{cont,m}}$) to the contact area of the bubble with the larger patch size prior to coalescence ($\mathrm{A_{cont,lp}}$). This reference is warranted by the observation that, during coalescence, the bubble with a larger contact area predominantly remains fixed to the surface, as seen in Fig. \ref{fig:SM_snapshot_sticking}. In the vast majority of cases ($>96\%$), the bubbles whose patch remains fixed is the larger one in the coalescing pair. Figure \ref{fig:8_cont_area} shows that for most cases $\mathrm{A_{cont,m}}/\mathrm{A_{cont,lp}}\approx 1$, implying that the patch size after coalescence equals the larger one prior to it. A few exceptions to this occur for $x \lessapprox 1.8$, where the patch size increases slightly relative to $\mathrm{A_{cont,lp}}$, especially when $R_{l,p}$ is comparatively small. Conversely, some cases with large $R_{l,p}$ also exhibit a decrease in patch size.

% \vspace{-1cm}

\section{\label{app:energetics}Energetics of jumping bubbles}
\setcounter{figure}{0}
\renewcommand{\thefigure}{B\arabic{figure}}

\setcounter{equation}{0}
\renewcommand{\theequation}{B\arabic{equation}}

\begin{figure}[h!]
	\centering
    \includegraphics[width=0.96\columnwidth]{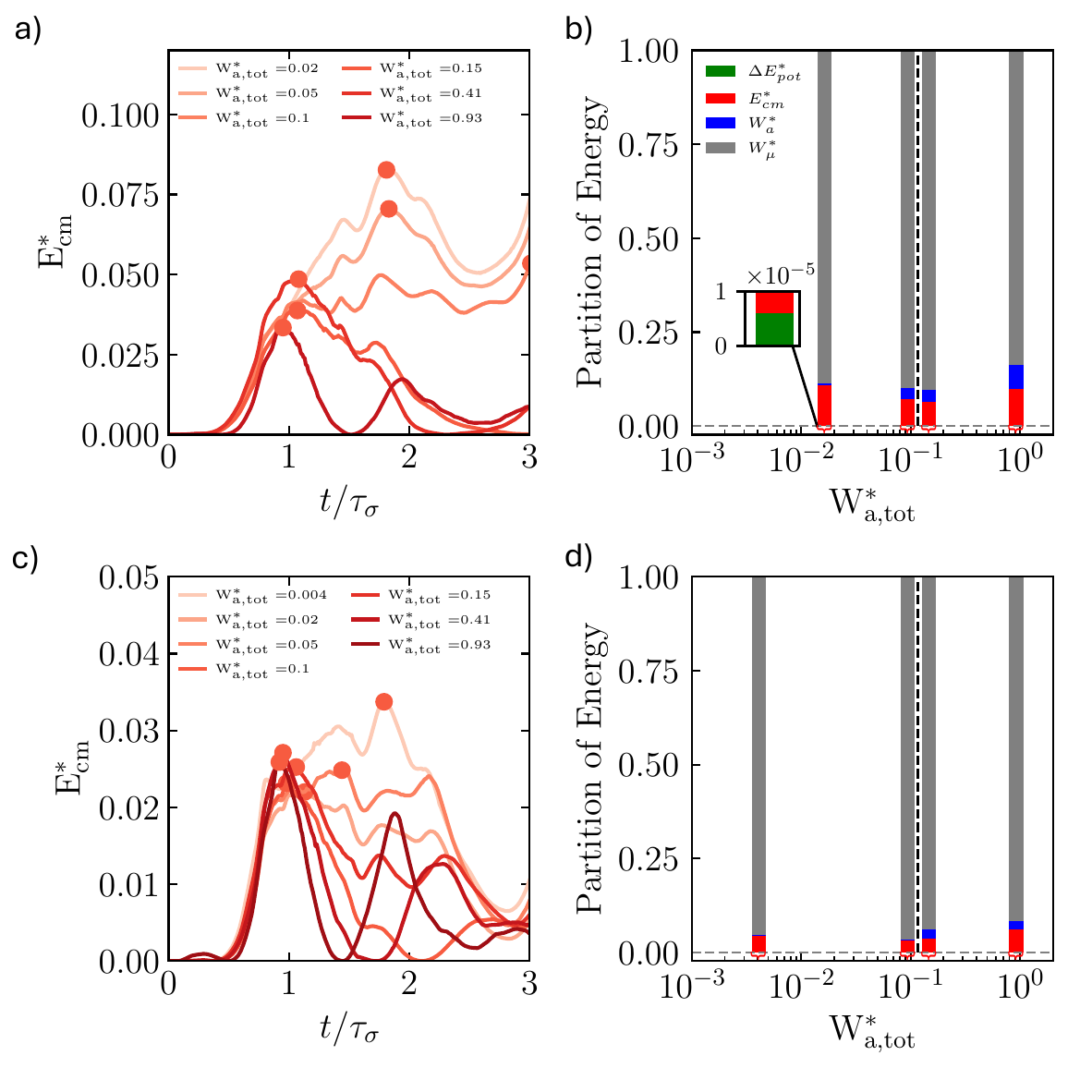}
	\caption{(a) Kinetic energy $E^*_{cm}$ (see Eq.~(\ref{eq:Ecm})) based on the center of mass velocity as a function of time for different $W^*_{a,tot}$. Here and in (b), $Bo=0.1$. The circles represent the times when $E^*_{cm}$ reaches the respective maximum. (b) The instantaneous energy partitions (all normalized by the instantaneous surface energy $E_s(t)-E_s(0)$) at the time when $E_{cm}^*$ reaches the maximum for different $W^*_{a,tot}$. The green part represents the change of potential energy, the red part represents the instantaneous $E_{cm}^*$, the blue part represents the instantaneous adhesion energy $W_a^*$, and the grey part represents the remaining dissipation energy $W_\mu^*$. The vertical dashed line shows the transition from jumping case to sticking case. (c)(d): the same plots as (a)(b) but for $Bo=0.001$.} 
\label{fig:SM_Energetics}
\end{figure}

In Fig.~\ref{fig:SM_Energetics}(a)(c), we plot the temporal evolution of the kinetic energy normalized by the total surface energy:
\begin{equation}
    E_{cm}^* = \frac{\frac{1}{2}C_M\rho_l\frac{4}{3}\pi R_m^3 V^2_{cm}}{4\pi\sigma(R_l^2+R_s^2-R_m^2)}=\frac{1}{3(2-2^{\frac{2}{3}})}C_MV^{*2}_{cm},
    \label{eq:Ecm}
\end{equation}
\newline
where $V_{cm}$ is the dimensional center of mass velocity, $V^*_{cm}$ is the non-dimensionlized center of mass velocity, and $C_M=0.8$ is the added mass coefficient. It shows that the kinetic energy accounts for only a small part of the total energy ($<10\%$) for both $Bo=0.1$ (Fig.~\ref{fig:SM_Energetics}(a)) and $Bo=0.001$ (Fig.~\ref{fig:SM_Energetics}(c)). We further plot the energy partitions, including kinetic energy, adhesion energy, potential energy and the viscous dissipation energy (taken as the remaining energy not accounted for by the other terms here), at the time when the kinetic energy reaches its respective maximum for $Bo=0.1$ (Fig.~\ref{fig:SM_Energetics}(b)) and $Bo=0.001$ (Fig.~\ref{fig:SM_Energetics}(d)). Note that the potential energy is too small to be visible in this partition plot (see blow-up in Fig.~\ref{fig:SM_Energetics}(b)). 
Analysis of the energy partitions reveals that, while most of the energy is dissipated, the adhesion energy can reach a magnitude comparable to $E_{cm}^*$, underscoring its significance in the problem.

\section{\label{app:derivation}Scaling relation for the energy dissipation}

\setcounter{figure}{0}
\renewcommand{\thefigure}{C\arabic{figure}}

\setcounter{equation}{0}
\renewcommand{\theequation}{C\arabic{equation}}

During bubble coalescence, the dissipated energy $W_\mu(t)$ up to time $t$ is given by the integral:
\begin{equation}
	\label{mainDiss}
	W_{\mu}(t) = \int_0^t \int_V \Phi(\boldsymbol{x},t')\, dV \, dt' ,
\end{equation}

\noindent where $t$ is time, $V$ is the volume over which dissipation occurs, and $\Phi$ is the viscous dissipation function given by
\begin{align}
	\Phi(\boldsymbol{x},t) = 2\mu\left(\boldsymbol{\mathcal{D}}(\boldsymbol{x},t)\boldsymbol{:\mathcal{D}}(\boldsymbol{x},t)\right).
\end{align}

\noindent Here, $\mu$ is the viscosity of the liquid and the deformation tensor $\boldsymbol{\mathcal{D}} = \left(\boldsymbol{\nabla v}(\boldsymbol{x},t) + \left(\boldsymbol{\nabla v}(\boldsymbol{x},t)\right)^T\right)/2$ is the symmetric part of the velocity gradient tensor. The coalescence-induced jumping phenomenon is primarily driven by capillary stresses. Following \cite{lv2021self,sanjay_chantelot_lohse_2023}, we can estimate the viscous dissipation $\Phi$ to leading order as
\begin{equation}
\Phi \sim \mu \frac{U_\sigma^2}{\lambda^2},
\end{equation}

\noindent where $U_{\sigma} \sim \sqrt{\sigma / \rho R_m}$ represents the inertio-capillary velocity. Here, $R_m$, the radius of the merged bubble, serves as the relevant length scale for the driving capillary stresses, and $\lambda$ denotes the length scale over which velocity gradients persist in the liquid. The continuity of velocity and viscous stresses across the gas-liquid interface dictates
\begin{align}
\mu \frac{U_\sigma}{\lambda} \sim \mu_\text{gas} \frac{U_\sigma}{R_m},
\label{eq:iface_cond}
\end{align}

\noindent implying that the unknown velocity gradients in the liquid can be approximated by those inside the merged bubble. We can then estimate the viscous dissipation function as $\Phi \sim \mu_\text{gas}U_\sigma^2/(R_m\lambda)$, which must be integrated across the dissipation volume  $V \sim \lambda R_m^2$ and over the inertio-capillary timescale of $\tau_\sigma \sim \sqrt{\rho R_m^3 / \sigma}$ to give the total viscous dissipation as
\begin{equation}
	\label{eq:vissDissFinal}
	W_{\mu} \sim \mu_\text{gas} \sqrt{{\sigma}/{\rho}} R_m^{3/2}.
\end{equation}

\noindent Normalizing Eq.~\ref{eq:vissDissFinal} with the released surface energy $\Delta{G}$ scale 
results in 
\begin{equation}
	\frac{W_{\mu}}{\Delta G} \sim \Bigl(\frac{\mu_\text{gas}}{\mu}\Bigr)\cdot Oh\cdot f(x).
    \label{eq:diss_appendC}
\end{equation}

\begin{figure}[h!]
	\centering	\includegraphics[width=0.8\columnwidth]{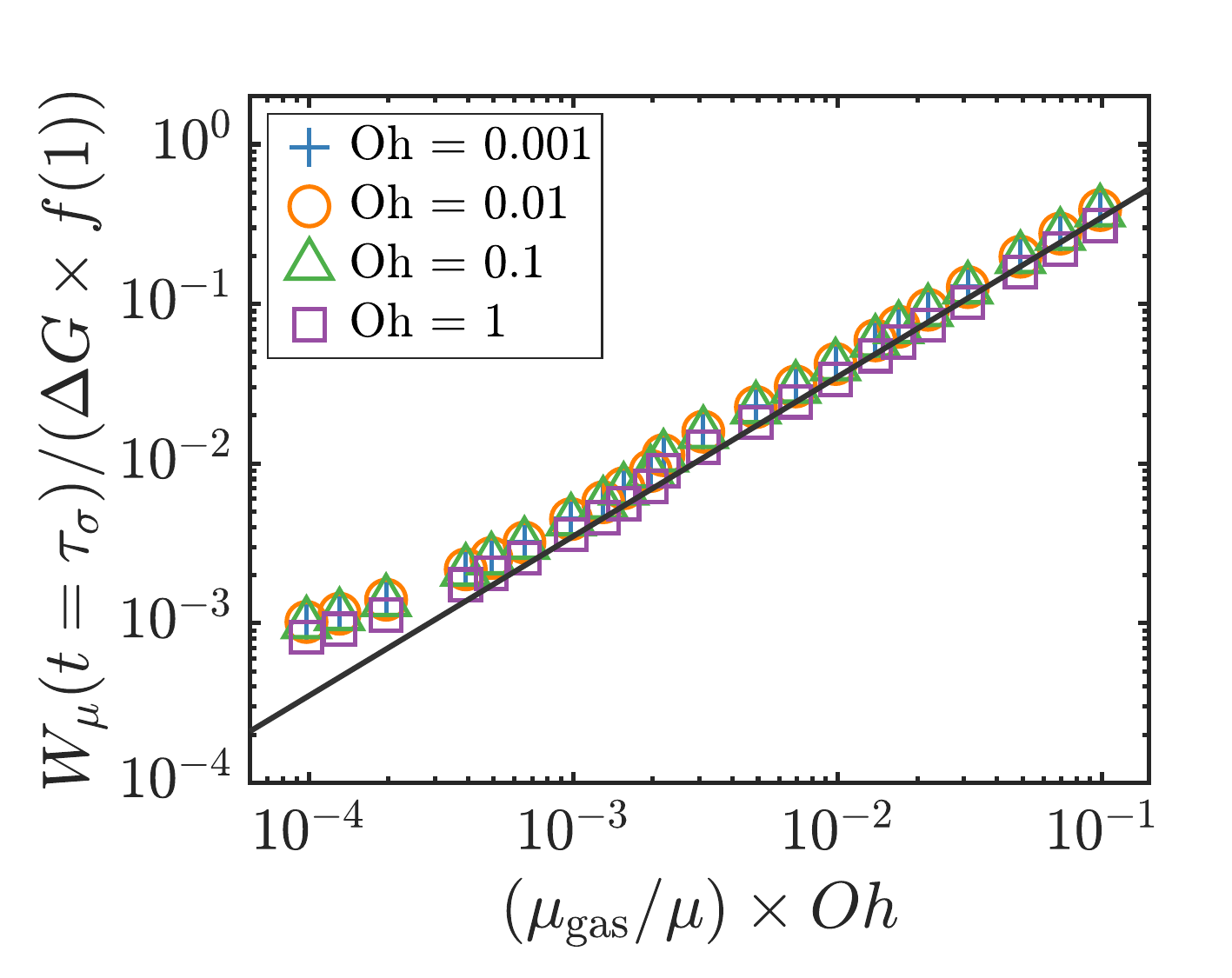}
	\caption{\rev{ Measured dissipation in the liquid over a period of $\tau_\sigma$ for coalescence of symmetric bubbles away from the wall with varying $Oh$ and $\mu_{gas}/\mu$. The line represents the scaling according to Eqn. \ref{eq:diss_appendC}.}}
	\label{fig:fx_change}
\end{figure}

\noindent 
\rev{This result is confirmed from simulations (see Fig. \ref{fig:fx_change})} and can be interpreted as an effective Ohnesorge number $(\mu_\text{gas}/\mu)\cdot Oh \cdot f(x)$ accounting for the viscosity ratio and size disparity. 

Note that the expression derived in \cite{lv2021self} for the same quantity amounts to setting $\lambda \approx R_m$, which yields $W_\mu \propto \mu$. However, this viscosity dependence is inconsistent with the experimental data in Ref. \cite{lv2021self}, in which it is argued that this discrepancy may be due to an (unaccounted) viscous damping effect that effectively cancels the $\mu$ dependence for $W_\mu$. The present formulation is consistent with the experimental data (see Fig. \ref{fig:4_Wa_LargeSmall}) without such an \emph{ad hoc} assumption. We emphasize that though $\mu_{gas}$ enters into Eq. (\ref{eq:vissDissFinal}) via the continuity condition Eq. (\ref{eq:iface_cond}), the relevant dissipation of energy occurs in the liquid.

\newpage

\bibliography{references}% Produces the bibliography via BibTeX.

\end{document}

\setcounter{figure}{0}
\setcounter{equation}{0}

\makeatletter 
\renewcommand{\thefigure}{S\@arabic\c@figure}
\renewcommand{\thetable}{S\@arabic\c@table}
\renewcommand{\thesection}{S\@arabic\c@section}
\renewcommand{\theequation}{S\@arabic\c@equation}

% \preprint{APS/123-QED}

\title{Supplementary material for: \\
	To jump or not to jump: Adhesion and viscous dissipation dictate the detachment of coalescing wall-attached bubbles}% Force line breaks with \\

\author{\c{C}ayan Demirk{\i}r}
% \email[]{c.demirkir@utwente.nl}
\affiliation{
	Physics of Fluids Department, Max Planck Center Twente for Complex Fluid Dynamics, and J. M. Burgers Center for Fluid Dynamics, University of Twente, P.O. Box 217, 7500AE Enschede, Netherlands
}

\author{Rui Yang}
%\email[]{c.demirkir@utwente.nl}
\affiliation{
	Physics of Fluids Department, Max Planck Center Twente for Complex Fluid Dynamics, and J. M. Burgers Center for Fluid Dynamics, University of Twente, P.O. Box 217, 7500AE Enschede, Netherlands
}

\author{Aleksandr Bashkatov}
%\email[]{c.demirkir@utwente.nl}
\affiliation{
	Physics of Fluids Department, Max Planck Center Twente for Complex Fluid Dynamics, and J. M. Burgers Center for Fluid Dynamics, University of Twente, P.O. Box 217, 7500AE Enschede, Netherlands
}

\author{Vatsal Sanjay}
%\email[]{c.demirkir@utwente.nl}
\affiliation{
	Physics of Fluids Department, Max Planck Center Twente for Complex Fluid Dynamics, and J. M. Burgers Center for Fluid Dynamics, University of Twente, P.O. Box 217, 7500AE Enschede, Netherlands
}

\author{Detlef Lohse}%
% \email[]{d.lohse@utwente.nl}
\affiliation{
	Physics of Fluids Department, Max Planck Center Twente for Complex Fluid Dynamics, and J. M. Burgers Center for Fluid Dynamics, University of Twente, P.O. Box 217, 7500AE Enschede, Netherlands
}
\affiliation{
	Max Planck Institute for Dynamics and Self-Organisation, Am Fassberg 17, 37077 G{\"o}ttingen, Germany
}

\author{Dominik Krug}
%\email[Contact author: ]{d.j.krug@utwente.nl}
\email[Contact author: ]{d.krug@aia.rwth-aachen.de}
\affiliation{
	Physics of Fluids Department, Max Planck Center Twente for Complex Fluid Dynamics, and J. M. Burgers Center for Fluid Dynamics, University of Twente, P.O. Box 217, 7500AE Enschede, Netherlands
}
\affiliation{
	Institute of Aerodynamics, RWTH Aachen University, Wüllnerstrasse 5a, 52062 Aachen, Germany
}

\date{\today}

% ------------------------------------------------------------------

\maketitle

\tableofcontents
\clearpage

\clearpage
\section{Sticking  bubbles: moving contact patch}
\label{supp:moving_patch}

Sticking bubble cases mostly result from the contact patch of one bubble remaining attached, while the other one gets peeled off (see Fig. \textcolor{blue}{A1}). However, in our experiments, we also observe instances of `moving contact patches', albeit less frequently. In these instances, both the size of the coalescing bubbles as well as the size of their contact patches are close to each other ($x \approx 1$ and $R_{cont,l}/R_{cont,s} \approx 1$), and the total dimensionless adhesion energy ($W^*_{a,tot}$) is always bigger than 0.2. Since the coalescence of two identical bubbles was simulated in this study, the experimental observation of bubbles sticking with a moving contact patch provides a relevant reference case and is therefore included here.

\begin{figure}[h!]
	\centering
	\includegraphics[width=0.95\columnwidth]{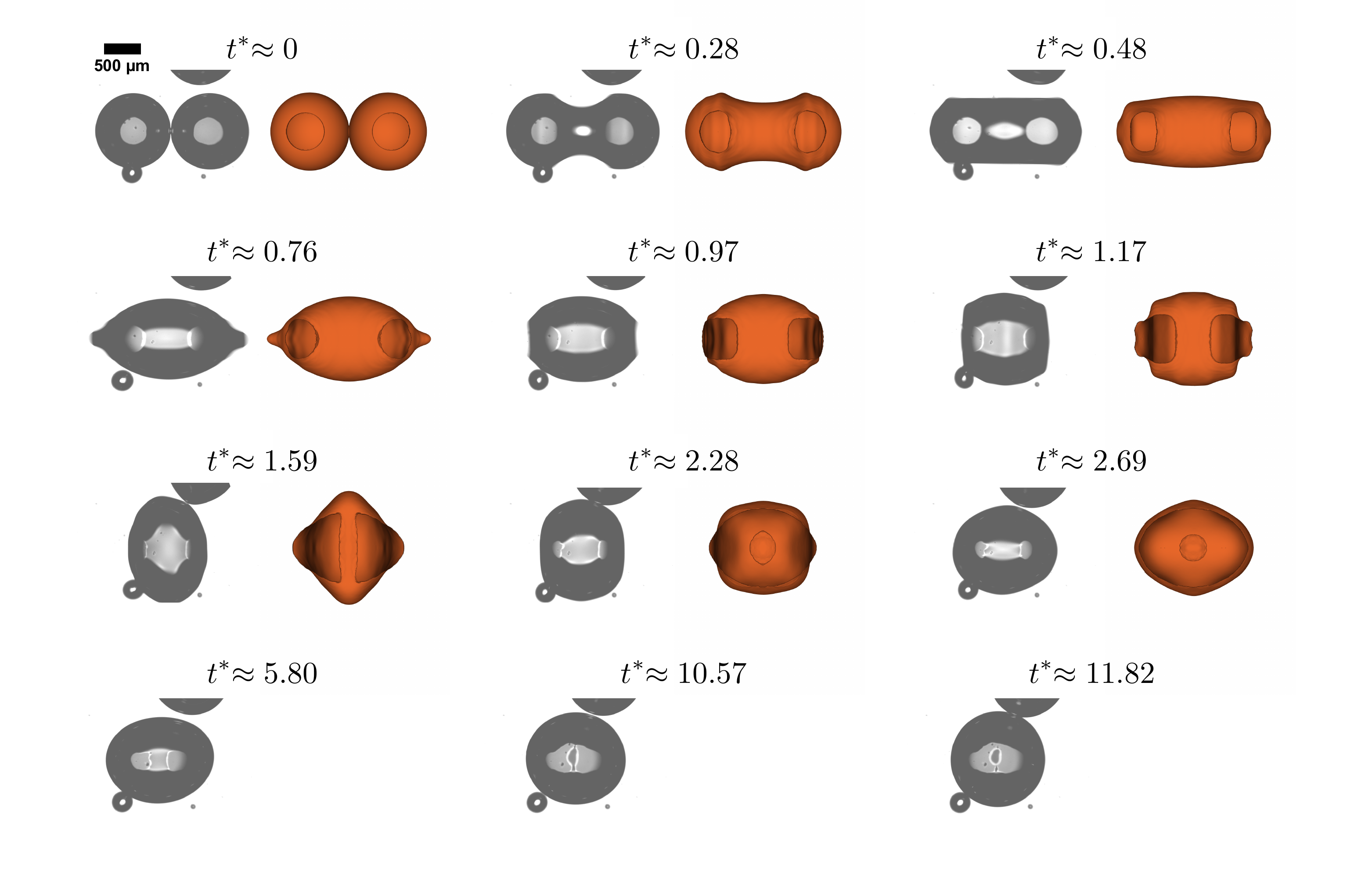}
	\caption{Shape and contact line evolution of two coalescing bubbles on a surface comparing experiment and simulations. Each pair shows: (left) experimental bottom view, and (right) bottom view from numerical simulation. In the experiments: $R_l \approx \SI{534}{\micro\meter},  R_s \approx \SI{525}{\micro\meter}$, $R_{\text{cont},l} = \SI{194}{\micro\meter}$ and $R_{\text{cont},s} = \SI{184}{\micro\meter}$. In the simulations: $R/R_{cont} = 0.365$. $Oh = 0.0052$, $Bo = 0.042$, {and $W^*_{a,\text{tot}} \approx 0.21$.\bb}. See also Movie 5 for further details.}
	\label{fig:SM_moving_snapshots}
\end{figure}

Figure \ref{fig:SM_moving_snapshots} presents experimental snapshots of an approximately symmetric coalescence event, alongside corresponding numerical simulations. In the experiments, the grey regions at the center of the bubbles correspond to the contact patches. Similarly, the edges of the contact patches are also discernible in the simulation images. The experiments and simulations show good agreement until $t^* \approx 1.2$. Beyond this point, differences emerge due to the different contact line dynamics in the simulation. Importantly, neither in the experiments nor in the simulation, the contact patches have merged at this point, where detachment already occurs in cases leading to departure (see Fig. \textcolor{blue}{1}). Therefore, it seems unlikely that a potential touchdown of the neck significantly influences the coalescence outcome as was suggested earlier \cite{iwata2022coalescing}. Merging of the patches occurs only in a later relaxation stage around $t^* = 2$ in the simulations, which are then stopped at  $t^* = 3$. In the experiments, this takes even longer until $t^* \approx 10$. In both cases a liquid drop becomes enclosed by the joining contact lines and remains within the contact patch. 
% \clearpage

\section{Buoyancy-driven detachment}
\label{supp:buo_driven}

When bubbles coalesce and subsequently detach, two primary modes of detachment are observed: coalescence-driven detachment, which occurs dynamically driven by the released surface energy, and buoyancy-induced detachment, which happens when the merged bubble's volume exceeds static buoyancy constraints. Our research aims to elucidate the conditions that trigger dynamic bubble detachment as a direct consequence of coalescence, distinguishing these cases from buoyancy-driven detachment.

As explained in a previous work \cite{demirkir2024life}, the detachment of a spreading bubble occurs when the contact angle of the bubble $\theta$ reaches to advancing contact angle $\theta_{adv}$, causing the contact line to begin shrinking. At this point, capillary forces can no longer keep the bubble adhered to the surface, and detachment occurs. At the onset of the contact line advancing, the force balance is approximately as follows:

\begin{equation}
	F_{\sigma} (\theta = \theta_{adv}) = F_b,
	\label{eq:force_bal_det}
\end{equation}
where, $F_\sigma$ and $F_b$ are the surface tension and buoyancy forces, respectively. Note that this assumes that other contributions, such as Marangoni forces, electric force, etc. can be neglected, which is appropriate for the present conditions \cite{demirkir2024life}. The buoyancy-driven cases are then detected by comparing the estimated detachment radius $R_{det,est}$, and the actual detachment radius $R_{det,act}$ of the merged bubble. From equation \ref{eq:force_bal_det},
\begin{equation}
	R_{det,est} = \left( \frac{3}{2} \frac{\sigma \cdot \sin{\theta_{adv}} \cdot R_{cont,l}}{\rho g} \right)^{1/3}.
	\label{eq:Rdet_est}
\end{equation}

We used larger bubble's contact radius $R_{cont,l}$ in our calculations since typically the smaller bubble's contact patch is swept by capillary waves first, then the merged bubble locates at the larger bubble's position regardless of whether coalescence results in detachment. Moreover, the advancing contact angle $\theta_{adv}$ was determined by the sessile drop experiments presented in Ref. \cite{demirkir2024life}, incorporating mean, upper, and lower limit values.

\begin{figure}[h!]
	\centering
	\includegraphics[width=0.75\textwidth]{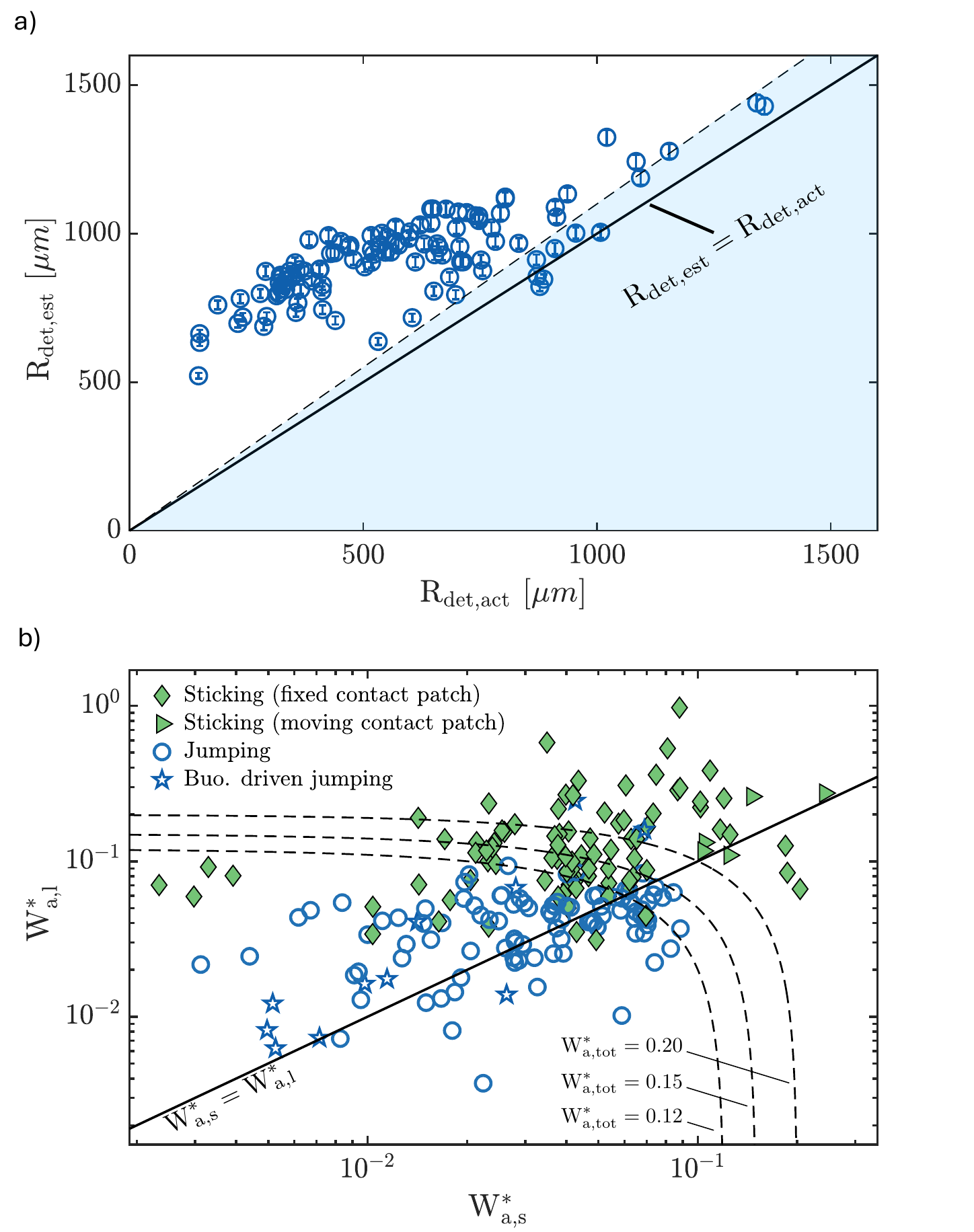}
	\caption{a) Estimated and detachment radii ($R_{det,est}$ and $R_{det,act}$, respectively) of the coalescence events result in jumping. Blue shaded area shows the buoyancy-driven detachment region, whereas the upper part of the dashed line is the cases jumped due to excess surface energy in coalescence. b) Dimensionless adhesion energy of smaller $W_{a,s}^*$ and larger bubbles $W_{a,l}^*$. The cases jumping occurs due to the excess surface energy shown by blue circle marker, while the blue star markers indicates the cases where merged bubble detached due to exceeding the buoyancy limits.}
	\label{fig:9_buo_det}
\end{figure}

The results of our analysis are presented in Figure \ref{fig:9_buo_det}a, which compares the estimated and actual detachment radii for coalescence cases resulting in bubble jumping. Most cases exhibit earlier jumping than predicted, with $R_{det,act}$ less than $R_{det,est}$. However, several cases exceeded buoyancy limitations, shown in the lower part of the diagonal where $R_{det,act}$ is greater than $R_{det,est}$. To account for the forces neglected in the force balance (equation \ref{eq:force_bal_det}) and other experimental errors, we left a 10\% margin above the diagonal line  ($R_{det,est}$ / $R_{det,act}$ $<$ 1.1) in our identification of buoyancy-driven cases. 
Cases that lie in the blue shaded region in Figure \ref{fig:9_buo_det}a were excluded from the results of this work.
For reference, the omitted cases are indicated with blue star markers in Figure \ref{fig:9_buo_det}, which shows a similar plot to Figure \textcolor{blue}{4} in the paper.

\clearpage
\section{Bubble size ratio and total adhesion energy relationship }
\label{supp:adh_Radratio}

\begin{figure}[h]
	\centering
	\includegraphics[width=0.75\textwidth]{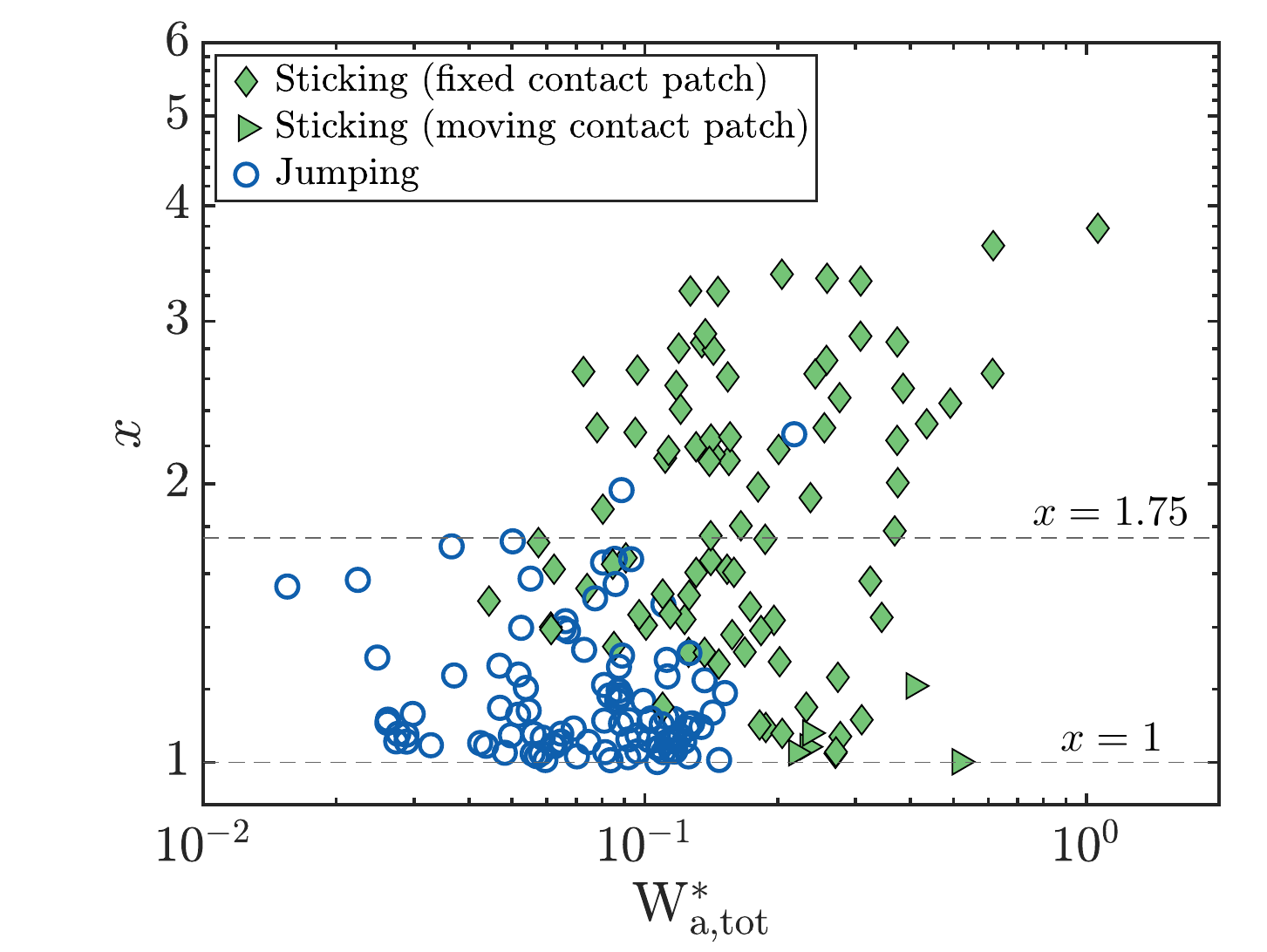}
	\caption{Radius ratio of the coalescing bubbles ($x$) with respect to the dimensionless total adhesion energy ($W_{a,tot}$). Green markers represent the sticking cases, while blue markers denote the jumping cases. The horizontal dashed lines serve as guides to indicate different radius ratios.}
\label{fig:SM_rad_rati_adhesion}
\end{figure}

The total adhesion energy of the coalescing bubbles ($\mathrm{W_{a,tot}}$) with respect to the radius ratio of the bubbles ($x$) is represented in Fig. \ref{fig:SM_rad_rati_adhesion}. It is seen that the jumping possibility of the bubbles significantly decreases as the $x$ increases. $x$ was found to be less than 1.75 for approximately 98\% of the jumping bubbles (blue circles). 

\vspace{-0.5cm}

%  \clearpage
\textcolor{black}{
\section{Distribution of the parameters investigated}}

\begin{figure}[h!]
	\centering
	\includegraphics[width=0.8\textwidth]{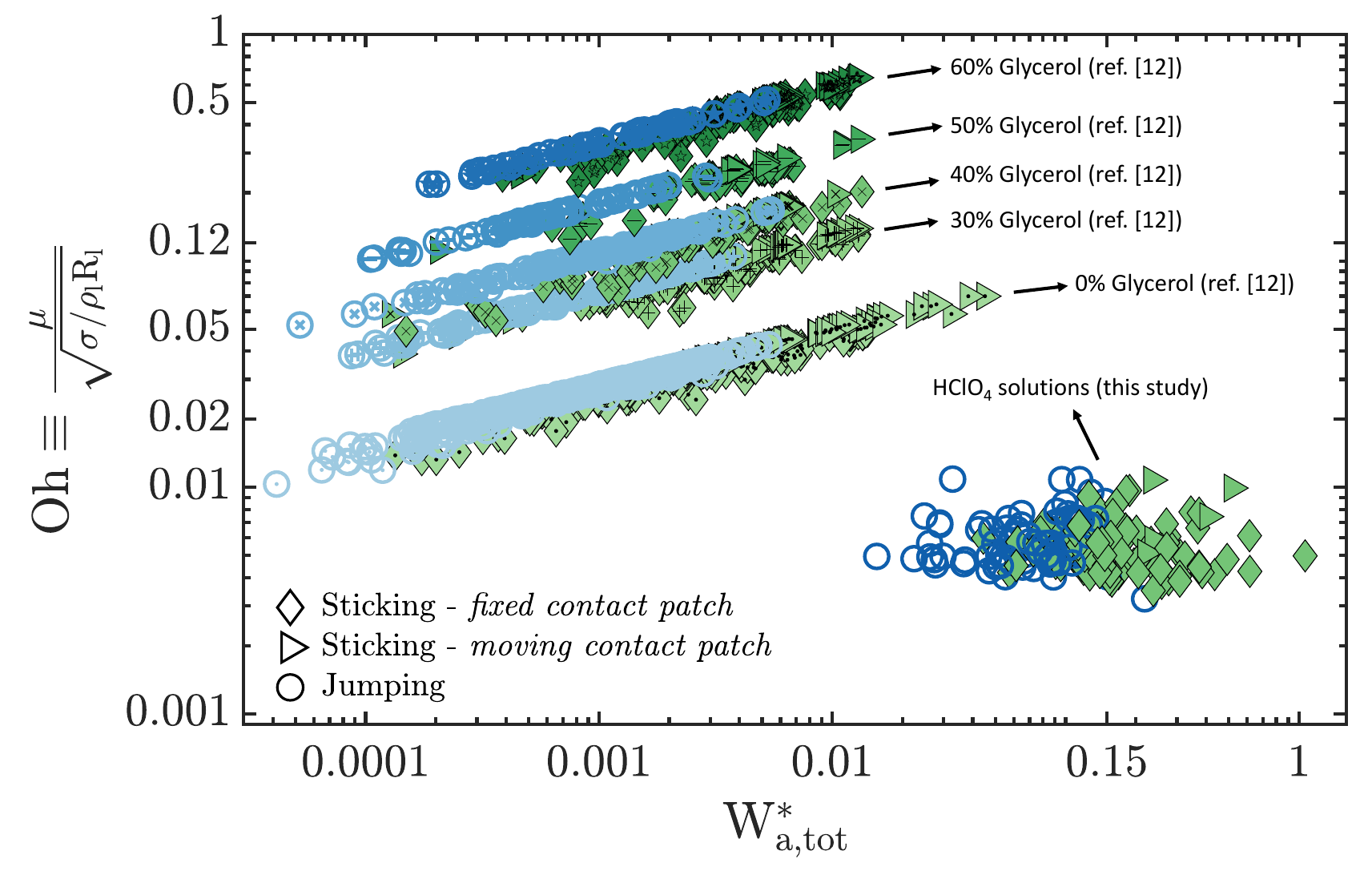}
	\caption{\textcolor{black}{Ohnesorge number (Oh) with respect to the dimensionless total adhesion energy. The electrolytes with varying glycerol concentrations also contain 30\% $\mathrm{H_2O_2}$ (see Ref. \cite{lv2021self} for more details). }}
	\label{fig:parameter_distr}
\end{figure}

\textcolor{black}{The parameters investigated in this study are presented via effective Ohnesorge number ($W^*_{\mu}$) in Fig. 4 of the main text. Although the data in Fig. 4 were obtained using electrolytes with different physical properties, this distinction may not be immediately apparent to the reader. To provide additional context, the viscosity ($\mu$) and surface tension ($\sigma$) values used in this study are listed in Table \ref{tab:param_distrib}. Furthermore, Fig. \ref{fig:parameter_distr} illustrates the distribution of Ohnesorge number (Oh) values as a function of the dimensionless total adhesion energy ($W^*_{a,tot}$).  It is noted that all experiments in this study, as well as those in Ref. \cite{lv2021self} (Ref. [14] in the main text), were conducted at room temperature ($\approx$ 20$^{\circ}$C).}

\begin{table}[ht]
\begin{tabular}{lll}
\toprule
& \shortstack[c]{\hspace{-1em}Viscosity ($\mu$) \\ {[mPa.s]}}  & \shortstack[c]{\hspace{-1.5em}Surface tension ($\sigma$)\hspace{-1em} \\ \hspace{1em}{[mN/m]}} \\ % Adjust the horizontal space (0.5em) here
\specialrule{\heavyrulewidth}{0pt}{1pt} % Increase height of the first horizontal line
Water & \hspace{2mm}1.00 \cite{korson1969viscosity} & 72.70 $\pm$ 0.09 \cite{demirkir2024life} \\
1 M HClO\textsubscript{4} (this study)&  \hspace{2mm}1.04 \cite{brickwedde1949properties} & 67.22 $\pm$ 0.10 \cite{demirkir2024life}\\
0.1 M HClO\textsubscript{4} + 0.5 M NaClO\textsubscript{4} (this study) &  \hspace{2mm}Not measured & 69.31 $\pm$ 0.07 \cite{demirkir2024life} \\
30\% H\textsubscript{2}O\textsubscript{2} + 0\% Glycerol (ref. \cite{lv2021self}) & \hspace{2mm}1.13 $\pm$ 0.03 & 71.04 $\pm$ 1.16 \\
30\% H\textsubscript{2}O\textsubscript{2} + 30\% Glycerol (ref. \cite{lv2021self}) & \hspace{2mm}2.91 $\pm$ 0.05 & 67.98 $\pm$ 1.74 \\
30\% H\textsubscript{2}O\textsubscript{2} + 40\% Glycerol (ref. \cite{lv2021self}) & \hspace{2mm}4.31 $\pm$ 0.09 & 67.02 $\pm$ 1.01 \\
30\% H\textsubscript{2}O\textsubscript{2} + 50\% Glycerol (ref. \cite{lv2021self}) & \hspace{2mm}7.06 $\pm$ 0.04 & 65.32 $\pm$ 0.84 \\
30\% H\textsubscript{2}O\textsubscript{2} + 60\% Glycerol (ref. \cite{lv2021self}) & \hspace{0mm}13.29 $\pm$ 0.17 & 64.45 $\pm$ 0.66 \\
%\specialrule{\heavyrulewidth}{2pt}{0pt} % Increase height of the second horizontal line
\bottomrule
\end{tabular}
\caption{\textcolor{black}{Viscosity ($\mu$) and surface tension ($\sigma$) at 20$^{\circ}$C of the electrolytes used in this study and those reported in Ref. \cite{lv2021self}.}}
\label{tab:param_distrib}
\end{table} 

\clearpage
\section{List of supplemental movies}

\begin{itemize}
	\item \textbf{Movie 1:} Jumping bubbles example. Corresponding to the snapshots shown in Fig. 1. (50,000 fps record, 10 fps play) 
	\item \textbf{Movie 2:} Sticking bubbles with fixed contact line example. Corresponding to the snapshots shown in Fig A1. (20,000 fps record, 10 fps play) 
	
    \item \textbf{Movie 3:} Sticking bubbles with moving contact line example. Corresponding to the snapshots shown in Fig S4. (10,000 fps record, 10 fps play)
    
    \item \textbf{Movie 4:} Jumping bubbles: experiments vs simulations. \newline (Experimental part: 50,000 fps record, 27 fps play) 
    
    \item \textbf{Movie 5:} Sticking bubbles with moving contact line: experiments vs simulations. (Experimental part: 10,000 fps record, 4.5 fps play)  
\end{itemize}
\clearpage
\bibliography{References}